\documentclass[aps,prl,reprint,floatfix,superscriptaddress,nobalancelastpage]{revtex4-2}

\usepackage{lipsum}
\usepackage{graphicx}
\usepackage{xcolor}
\usepackage{amsmath}
\usepackage{amsfonts}
\usepackage{todonotes}
\usepackage{lineno}
\usepackage{accents}
\usepackage{booktabs}
\newlength{\dhatheight}
\newcommand{\doublehat}[1]{%
    \settoheight{\dhatheight}{\ensuremath{\hat{#1}}}%
    \addtolength{\dhatheight}{-0.35ex}%
    \hat{\vphantom{\rule{1pt}{\dhatheight}}%
    \smash{\hat{#1}}}}

\newcommand{\musurface}{\ensuremath{\mu^{\mathrm{bkg}}_{\mathrm{surf.}}}}
\newcommand{\muthermal}{\ensuremath{\mu^{\mathrm{bkg}}_{\mathrm{therm.}}}}
\newcommand{\mutirbkg}{\ensuremath{\mu^{\mathrm{bkg}}_{\mathrm{TIR}}}}
\newcommand{\musr}{\ensuremath{\mu_{\mathrm{pass.}}}}
\newcommand{\tautir}{\ensuremath{\tau_{\mathrm{TIR}}}}
\newcommand{\mutir}{\ensuremath{\mu_{\mathrm{TIR}}}}
\newcommand{\nuvec}{\ensuremath{\boldsymbol{\nu}}}
\newcommand{\alphavec}{\ensuremath{\boldsymbol{\alpha}}}
\newcommand{\musig}{\ensuremath{\mu_{\mathrm{sig}}}}
\newcommand{\musigest}{\ensuremath{\hat{\mu}_{\mathrm{sig}}}}
\newcommand{\Po}{\ensuremath{\mathrm{Po}}}

\newcommand{\imp}{\ensuremath{\mathcal{P}}}
\newcommand{\ipred}{\ensuremath{\mathcal{I}}}
\newcommand{\ipredn}{\ensuremath{{\ipred{}}}}

\newcommand{\ipredlf}{\ensuremath{\ipred_{\mathrm{LF}}}}
\newcommand{\ipredhf}{\ensuremath{\ipred_{\mathrm{HF}}}}

\newcommand{\om}{\omega}
\newcommand{\avg}[1]{\langle#1\rangle}
\newcommand{\abs}[1]{\left|#1\right|}

\newcommand{\clust}{\ensuremath{\overline{c}}}
\newcommand{\isolated}{\ensuremath{c}}
\newcommand{\lowimp}{\ensuremath{\overline{\imp{}}}}
\newcommand{\clustlowimp}{\ensuremath{c\overline{\imp{}}}}
\newcommand{\highimp}{\ensuremath{\imp{}}}
\newcommand{\nosun}{\ensuremath{n}}
\newcommand{\tauclust}{\ensuremath{\tau_c}}
\newcommand{\tauimp}{\ensuremath{\tau_{\imp{}}}}

\newcommand{\ipredmax}{\ensuremath{\mathcal{I}}_{\mathrm{max}}}

\newcommand{\Lsurf}{\ensuremath{L_{\mathrm{surf.}}}}

\newcommand{\muthermalbkg}{\ensuremath{\mu^{\mathrm{bkg}}_{\mathrm{therm.}}}}
\newcommand{\musurfacebkg}{\ensuremath{\mu^{\mathrm{bkg}}_{\mathrm{surf.}}}}

\newcommand{\mulowimp}{\ensuremath{\mu_{\lowimp{}}}}
\newcommand{\muhighimp}{\ensuremath{\mu_{\highimp{}}}}
\newcommand{\muclustlowimp}{\ensuremath{\mu_{\clustlowimp{}}}}
\newcommand{\muunclust}{\ensuremath{\mu_{\overline{c}}}}
\newcommand{\muclust}{\ensuremath{\mu_{c}}}
\newcommand{\munight}{\ensuremath{\mu_{n}}}
\newcommand{\thetaclus}{\ensuremath{\theta_\mathrm{clust.}}}

\newcommand{\flin}{\ensuremath{f_{\mathrm{lin}}}}
\newcommand{\fRMS}{\ensuremath{f_{\mathrm{rms}}}}

\newcommand{\corsika}{\mbox{CORSIK\kern -0.05em A 8}}
\newcommand{\eisvogel}{Eisvogel}

\newcommand{\sibyll}{\textsc{sibyll} 2.3d}
\newcommand{\fluka}{FLUKA 2024.1}
\newcommand{\proposal}{PROPOSAL 7.6.2}
 
\begin{document}

\date{\today}
\title{Observation of In-ice Askaryan Radiation from High-Energy Cosmic Rays}
\newcommand{\atUC}{\affiliation{Dept.~of Physics, Dept.~of Astronomy and Astrophysics, Enrico Fermi Institute, Kavli Institute for Cosmological Physics, University of Chicago, Chicago, IL 60637}}
\newcommand{\atKU}{\affiliation{Dept.~of Physics and Astronomy, University of Kansas, Lawrence, KS 66045}}
\newcommand{\atOSU}{\affiliation{Dept.~of Physics, Center for Cosmology and AstroParticle Physics, The Ohio State University, Columbus, OH 43210}}
\newcommand{\atChiba}{\affiliation{Dept.~of Physics, Chiba University, Chiba, Japan}}
\newcommand{\atUW}{\affiliation{Dept.~of Physics, University of Wisconsin-Madison, Madison,  WI 53706}}
\newcommand{\atNTU}{\affiliation{Dept.~of Physics, Graduate Institute of Astrophysics, Leung Center for Cosmology and Particle Astrophysics, National Taiwan University, Taipei, Taiwan}}
\newcommand{\atULB}{\affiliation{Universite Libre de Bruxelles, Science Faculty CP230, B-1050 Brussels, Belgium}}
\newcommand{\atUMD}{\affiliation{Dept.~of Physics, University of Maryland, College Park, MD 20742}}
\newcommand{\atUCL}{\affiliation{Dept.~of Physics and Astronomy, University College London, London, United Kingdom}}
\newcommand{\atPSUigc}{\affiliation{Center for Multi-Messenger Astrophysics, Institute for Gravitation and the Cosmos, Pennsylvania State University, University Park, PA 16802}}
\newcommand{\atPSUphys}{\affiliation{Dept.~of Physics, Pennsylvania State University, University Park, PA 16802}}
\newcommand{\atPSUast}{\affiliation{Dept.~of Astronomy and Astrophysics, Pennsylvania State University, University Park, PA 16802}}
\newcommand{\atVUB}{\affiliation{Vrije Universiteit Brussel, Brussels, Belgium}}
\newcommand{\atUNL}{\affiliation{Dept.~of Physics and Astronomy, University of Nebraska, Lincoln, Nebraska 68588}}
\newcommand{\atWhittier}{\affiliation{Dept.~Physics and Astronomy, Whittier College, Whittier, CA 90602}}
\newcommand{\atUD}{\affiliation{Dept.~of Physics, University of Delaware, Newark, DE 19716}}
\newcommand{\atNUU}{\affiliation{Dept.~of Energy Engineering, National United University, Miaoli, Taiwan}}
\newcommand{\atNPU}{\affiliation{Dept.~of Applied Physics, National Pingtung University, Pingtung City, Pingtung County 900393, Taiwan}}
\newcommand{\atDenison}{\affiliation{Dept.~of Physics and Astronomy, Denison University, Granville, Ohio 43023}}
\newcommand{\atNDL}{\affiliation{National Nano Device Laboratories, Hsinchu 300, Taiwan}}

 \author{N.~Alden}\atUC
 \author{S.~Ali}\atKU
 \author{P.~Allison}\atOSU
 \author{S.~Archambault}\atChiba
 \author{J.J.~Beatty}\atOSU
 \author{D.Z.~Besson}\atKU
 \author{A.~Bishop}\atUW
 \author{P.~Chen}\atNTU
 \author{Y.C.~Chen}\atNTU
 \author{Y.-C.~Chen}\atNTU
 \author{S.~Chiche}\atULB
 \author{B.A.~Clark}\atUMD
 \author{A.~Connolly}\atOSU
 \author{K.~Couberly}\atKU
 \author{L.~Cremonesi}\atUCL
 \author{A.~Cummings}\atPSUigc\atPSUphys\atPSUast
 \author{P.~Dasgupta}\atOSU
 \author{R.~Debolt}\atOSU
 \author{S.~de~Kockere}\atVUB
 \author{K.D.~de~Vries}\atVUB
 \author{C.~Deaconu}\atUC
 \author{M.A.~DuVernois}\atUW
 \author{J.~Flaherty}\atOSU
 \author{E.~Friedman}\atUMD
 \author{R.~Gaior}\atChiba
 \author{P.~Giri}\atUNL
 \author{J.~Hanson}\atWhittier
 \author{N.~Harty}\atUD
 \author{K.D.~Hoffman}\atUMD
 \author{M.-H.~Huang}\atNTU\atNUU
 \author{K.~Hughes}\atOSU
 \author{A.~Ishihara}\atChiba
 \author{A.~Karle}\atUW
 \author{J.L.~Kelley}\atUW
 \author{K.-C.~Kim}\atUMD
 \author{M.-C.~Kim}\atChiba
 \author{I.~Kravchenko}\atUNL
 \author{R.~Krebs}\atPSUigc\atPSUphys
 \author{C.Y.~Kuo}\atNTU
 \author{K.~Kurusu}\atChiba
 \author{U.A.~Latif}\atVUB
 \author{C.H.~Liu}\atUNL
 \author{T.C.~Liu}\atNTU\atNPU
 \author{W.~Luszczak}\atOSU
 \author{A.~Machtay}\atOSU
 \author{K.~Mase}\atChiba
 \author{M.S.~Muzio}\atUW\atPSUigc\atPSUphys\atPSUast
 \author{J.~Nam}\atNTU
 \author{R.J.~Nichol}\atUCL
 \author{A.~Novikov}\atUD
 \author{A.~Nozdrina}\atOSU
 \author{E.~Oberla}\atUC
 \author{C.W.~Pai}\atNTU
 \author{Y.~Pan}\atUD
 \author{C.~Pfendner}\atDenison
 \author{N.~Punsuebsay}\atUD
 \author{J.~Roth}\atUD
 \author{A.~Salcedo-Gomez}\atOSU
 \author{D.~Seckel}\atUD
 \author{M.F.H.~Seikh}\atKU
 \author{Y.-S.~Shiao}\atNTU\atNDL
 \author{J. Stethem}\atOSU
 \author{S.C.~Su}\atNTU
 \author{S.~Toscano}\atULB
 \author{J.~Torres}\atOSU
 \author{J.~Touart}\atUMD
 \author{N.~van~Eijndhoven}\atVUB
 \author{A.~Vieregg}\atUC
 \author{M.~Vilarino~Fostier}\atULB
 \author{M.-Z.~Wang}\atNTU
 \author{S.-H.~Wang}\atNTU
 \author{P.~Windischhofer}\atUC
 \author{S.A.~Wissel}\atPSUigc\atPSUphys\atPSUast
 \author{C.~Xie}\atUCL
 \author{S.~Yoshida}\atChiba
 \author{R.~Young}\atKU
\collaboration{ARA Collaboration}\noaffiliation

\begin{abstract}
\noindent
We present the first experimental evidence for in-ice Askaryan radiation---coherent charge-excess radio emission---from 
high-energy particle cascades developing in the Antarctic ice sheet.
In 208 days of data recorded with the phased array instrument of the Askaryan Radio Array,
a previous analysis has incidentally identified 13 events with
impulsive radio frequency signals originating from below the ice surface.
We here present a detailed reanalysis of these events.
The observed event rate, radiation arrival directions, signal shape, spectral content, and electric field polarization
are consistent with in-ice Askaryan radiation from cosmic ray air shower cores impacting the ice sheet.
For the brightest events, the angular radiation pattern favors an extended cascadelike emitter over
a pointlike source.
An origin from the geomagnetic separation of charges in cosmic ray air showers is disfavored by the arrival directions and polarization.
Considering the arrival angles, timing properties, and impulsive nature of the passing events,
the event rate is inconsistent with the 
estimation of the combined background from thermal noise events and on-surface events at the level of $5.1\,\sigma$.
\end{abstract}

\maketitle

\mbox{}
\clearpage
\textsl{Introduction}---%
Over the last decade, 
optical Cherenkov detectors such as IceCube~\cite{IceCube_instrument} and 
KM3NeT~\cite{KM3NeT_instrument} have been very successful in probing extreme astrophysical environments
through observations of the cosmic neutrino flux~\cite{icecube-ngc-1068, icecube-txs-0506, icecube-txs-mma} below $\mathcal{O}(100\,\mathrm{PeV})$.
Extending these observations to EeV energies promises to help constrain the yet-unknown sources and 
composition of ultrahigh-energy cosmic rays~\cite{uhecr-constraints}.
However, the small expected flux~\cite{GZK_spectrum, more_GZK_calcs} requires instrumented volumes of 
$\mathcal{O}(100\,\mathrm{km}^3)$ water equivalent, which are difficult to realize with 
optical detectors.

To this end, the radio detection technique has emerged as a scalable experimental method to achieve the required
sensitivity, using glacial ice in the polar regions as the target medium.
Contemporary radio neutrino detectors target the coherent, Cherenkov-like electromagnetic radiation in the 
100\,MHz--1\,GHz band that is emitted by a neutrino-induced in-ice particle cascade and can propagate over several kilometers
through the radio-transparent ice~\cite{atten_length}.
First predicted by Askaryan in 1962~\cite{askaryan_1, askaryan_2}, this radiation has its origin in the net-negative charge generated
in the moving shower front as Compton, Bhabha, and M{\o}ller scattering draws electrons from the surrounding material
into the shower and positrons continuously annihilate~\cite{ZHSfreq}.
This radio frequency (rf) signature was first observed and studied in a series of experiments at particle accelerators in the 
2000s~\cite{sand_askaryan, salt_askaryan, polyethylene_askaryan, ice_askaryan},
which identified the Askaryan effect as the relevant emission mechanism for cascades in dense dielectrics.
Outside the laboratory, in-air Askaryan emission from cosmic-ray-induced air showers was first experimentally 
identified in 2014~\cite{Auger_air_askaryan, LOFAR_air_askaryan, CODALEMA_air_askaryan}.
However, in this case it only accounts for $\approx10\%$ of the total electric field 
amplitude~\cite{LOFAR_air_askaryan, CODALEMA_air_askaryan}, with emission from 
geomagnetic charge separation being dominant~\cite{original_geomagnetic, geomag_theory_1, geomag_theory_2}.
Supported by these results, a number of in-ice radio neutrino detectors have been deployed in 
Antarctica~\cite{arianna-search, RICE, ARA_instrument_paper, PA_instrument_paper} and Greenland~\cite{RNO-G_instrument_paper}.

Recently, theoretical studies have shown that cosmic rays (CRs) can also act as a source of in-ice Askaryan radiation for these 
observatories: the dense core of a near-vertical CR air shower can reach the surface of high-altitude polar plateaus 
and initiate an in-ice particle cascade~\cite{coleman2024iniceaskaryanemissionair, FAERIE} which
generates downgoing Askaryan radiation in the top 5--10\,m of the ice sheet~\cite{Original_faerie, FAERIE, De_Kockere_2024, shower_sim_thesis}.
In this Letter, we use data from the Askaryan Radio Array (ARA) to establish the first observation of this process.
This marks the first detection of a cosmic particle using its in-ice Askaryan emission,
which constitutes a new detection method for CRs and is an important benchmark for radio detectors that aim to make the 
first measurements of ultrahigh-energy neutrinos.

\begin{figure}[tb]
  \includegraphics[width=\columnwidth]{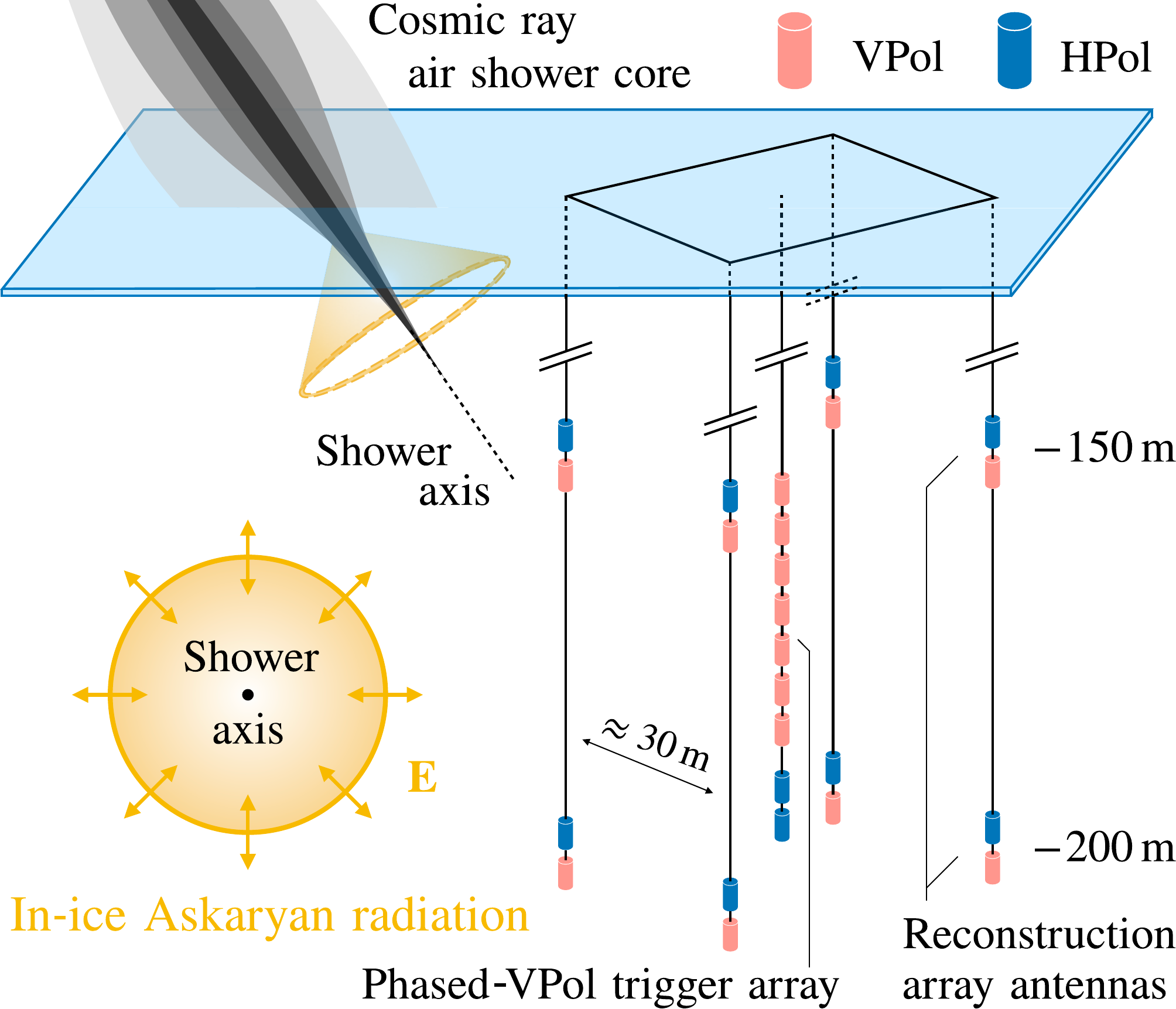}
  \caption{
	  The dense core of a CR air shower (black) emits Askaryan radiation in the near-surface ice, showing 
	  a characteristic radial electric field polarization with respect to the shower axis (bottom left).
	  The radiation is beamed around the Cherenkov angle, represented by the yellow cone.
	  A5 with the phased array is shown on the right.
  }
  \label{fig:detector}
\end{figure}

\textsl{Detector and event selection}---%
The ARA radio neutrino detector is located on the Antarctic Plateau, at $\approx$~2800\,m elevation, near the South Pole 
(at an atmospheric depth of $\approx 700\,\mathrm{g/cm}^2$).
ARA consists of five independent detector stations, spaced $\approx$ 2\,km apart.
Each station comprises eight vertically polarized antennas (VPol) and eight horizontally polarized antennas (HPol), 
with respective bandwidths of 150\,MHz--850\,MHz and 200\,MHz--850\,MHz,
deployed as strings in four air-filled boreholes at 150\,--200\,m depth~\cite{ARA_instrument_paper}.
ARA Station 5 (A5), used for this Letter and shown in Fig.~\ref{fig:detector}, is further equipped with
a compact array of seven VPol antennas and two HPol antennas in a central string.
They are used by a dedicated data-acquisition (DAQ) system to form a phased-VPol rf trigger \cite{PA_instrument_paper},
lowering the trigger threshold relative to the thermal electromagnetic radiation (``thermal noise'') emitted by the ice.
The antennas on the four outer strings (referred to as the reconstruction array) are used to find the azimuthal location of radiation sources. 
A nanosecond-pulsed calibration transmitter is located in a dedicated borehole $\approx$~50\,m from the center of A5.

In the following, we examine events that have previously been identified in a neutrino 
search \cite{ara-lowthresh-analysis}, conducted on 208 days of data recorded with the phased-VPol trigger 
in the period from March to November 2019.
Additional details and supporting plots for our analysis can be found in the End Matter and Supplemental Material
to this Letter \cite{supmat-arxiv}.
\InputIfFileExists{nocite}{}{}
The event selection uses a multivariate linear discriminant (LDA) trained to identify Askaryan emission from
simulated neutrino events.
The neutrino search region is defined, among other cuts, by requiring the radiation arrival zenith angle
\footnote{The zenith angle is measured between the vertical direction and the arrival direction of the 
radiation wavefront.}
measured at the phased array, $\theta$, to be greater than $57^\circ$.
The angle $\theta$ is reconstructed using an interferometric technique from the signals 
measured by the phased-VPol receivers, following Ref.~\cite{ara-lowthresh-analysis}.
Following unblinding, 46 events were observed with $\theta < 57^\circ$, adjacent to the neutrino search region.
\mbox{Here, we} study the events in the zenith region \mbox{$38^\circ \leq \theta \leq 57^\circ$}, where we 
expect signals from sources in the top few meters of the ice.
Radiation from distant above-surface sources, including anthropogenic emissions and geomagnetic radiation 
from CR air showers, is refracted at the air-ice interface and arrives at the phased array with $\theta < 34.5^\circ$, assuming a 
flat surface.
The limiting angle corresponds to total internal reflection (TIR) at the surface.
Our choice of zenith region, therefore, suppresses such components and selects a low-background event sample.

A total of 13 events are observed in the targeted region.
They are impulsive, occur at a uniform rate of $22.9^{+8.2}_{-6.2}\,\mathrm{yr}^{-1}$, and show no preferred azimuthal 
arrival direction.
All other rf triggers within 2\,h of each event are consistent with thermal noise.
The event times modulo day of the week, UTC hour, minute, and second are consistent with uniform distributions.
Twelve events occur during polar night (March~21 to September~21), consistent with the live time fraction, when human activity at and near
South Pole Station ($\approx$~5\,km 
from A5) is lowest.
The distribution of wind speeds at the event times is consistent with the wind-speed distribution for the full dataset,
indicating no preference for emissions from triboelectric surface discharges during high-wind periods~\cite{triboelectric}.

\textsl{Background estimate}---%
To compare the observed event yield with expected nonphysics processes, we construct a postunblinding background estimate
for three classes of backgrounds.
Each is assumed to be a Poisson process over the analyzed live time.
First, the background rate from thermal noise triggers is estimated at $0.14^{+0.05}_{-0.03}$ events
using the procedure of Ref.~\cite{ara-lowthresh-analysis}.
Second, the background rate from distant near-horizon sources leaking into the zenith region of interest is extrapolated from
a zenith control region including the TIR angle.
It is conservatively estimated at $0.15 \pm {0.04} \, \mathrm{(stat.)}\,{}^{+0.07}_{-0.03}\,\mathrm{(syst.)}$ events; 
an alternative modeling approach reduces this rate by a factor of $\approx$ 3.
The statistical and systematic components arise, respectively, from the event yield in the control region and
from uncertainties on the zenith angle reconstruction.
Third, signals from on-surface sources within $\approx$ 250\,m of A5 can enter the zenith region of interest through
evanescent coupling with the ice~\cite{Jackson_EM}.
Using control samples enriched in known on-surface activity, we find that such processes typically produce time-clustered triggers, of 
which only a small fraction ($\lesssim$10\%) are as impulsive as the passing events.
The impulsivity is quantified by the ratio of maximum instantaneous signal power to mean signal power~%
\footnote{
In this calculation, signal power is given by the magnitude square of the analytic signal.
To reduce the correlation of impulsivity with signal-to-noise ratio, this calculation is only applied to the 
50\% of the recorded trace closest to the peak.}.
Assuming these samples to be representative, we find
a Feldman-Cousins upper limit on the rate of 0.12 events at 95\% confidence level for the population
of time-unclustered on-surface background events with impulsivity similar to the passing events.
Modeling systematic uncertainties for on-surface sources contribute $\approx$ 25\% of this upper limit,
with the remainder due to the statistical uncertainty.

The observed event yield represents an excess over the combined background estimate at a
significance of $5.1\,\sigma$, assessed from the distribution of the profile-likelihood ratio test statistic~\cite{pdg-statistics} 
under the background-only hypothesis.
If impulsivity information is not explicitly used in the background estimate,
the significance is $3.5\,\sigma$.

\textsl{Askaryan event rate}---%
To estimate the Askaryan event rate expected from impacting air shower cores, we follow the Monte Carlo approach of
Ref.~\cite{coleman2024iniceaskaryanemissionair}.
We use a parametrized Askaryan emission model \cite{ARZ} for neutrino-induced in-ice cascades to efficiently 
simulate a large number of impacting shower-core-like events and weight them by the CR flux at the ice surface.
This approach incorporates a realistic ice refractive index profile 
and reproduces the Cherenkov-like angular emission pattern and radial polarization, 
thereby modeling the radiation arrival direction and polarization observables.
However, the simulated neutrino-induced cascades do not accurately predict the emission strengths of CR-induced 
\mbox{in-ice} cascades due to unaccounted-for differences in the charge-excess profile.
We approximate this effect by an amplitude scale factor  
determined through comparisons with microscopic simulations of impacting 
shower cores~\cite{coleman2024iniceaskaryanemissionair}.
As a result, absolute yield estimates carry sizable uncertainties, and signal shape observables are not
reliably predicted.
Therefore, we do not attempt to use the simulated LDA score but equivalently model the analysis selection threshold
in terms of the signal-to-noise ratio (SNR) averaged over all phased array antennas,
where the per-antenna SNR is defined as the 
ratio of the peak-to-peak signal amplitude and twice the thermal noise rms amplitude.
We use a cut value of SNR $>$ 4.6,
estimated from impulsive near-threshold events.
We predict a rate of 8--34\,yr$^{-1}$, 
consistent with the observed rate and prior calculations \cite{coleman2024iniceaskaryanemissionair}.
Accounting for energy losses in the atmosphere, the primary CR energy of the passing events is estimated to be of order $10^{17}$\,eV,
cf.~Fig.~S14 in Supplemental Material \cite{supmat-arxiv}.

\begin{figure}
  \includegraphics[width=\columnwidth]{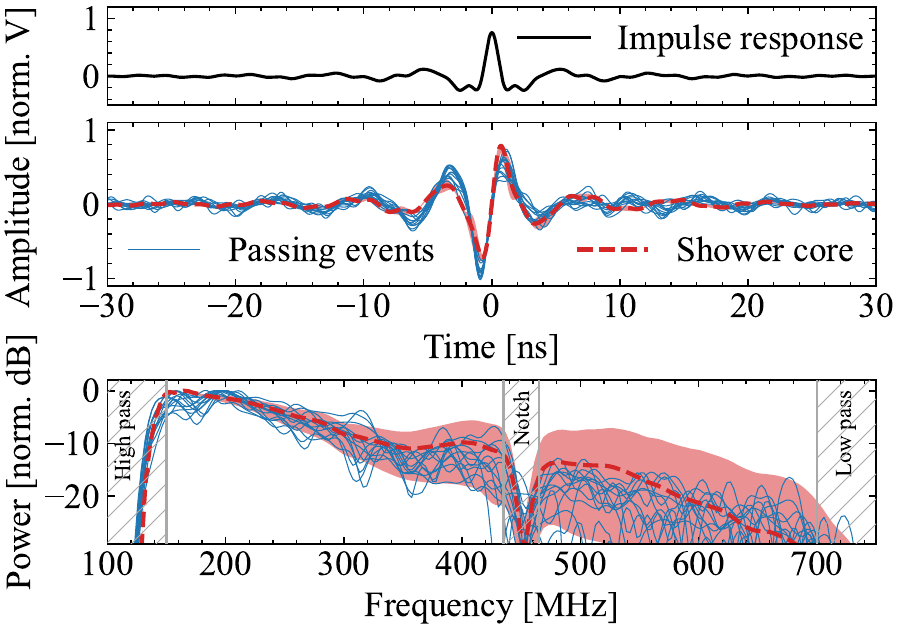}
  \caption{    
	Top, center: impulse response of the instrument (black) and VPol waveforms of the passing events (blue),
	with the instrumental phase response removed and normalized by the peak-to-peak amplitude.
	Bottom: relative power density spectra normalized by the maximum power density.
	Simulated signals are shown in red.
	The envelope indicates a range of $\approx$ $3^\circ$ around the approximate in-ice Cherenkov angle, 
	representative of events passing the selection.
  }
  \label{fig:signals}
\end{figure}

\textsl{Signal shape}---%
We next examine the VPol waveforms of all passing events.
They are displayed in Fig.~\ref{fig:signals} with the instrumental phase 
response removed and the time-aligned waveforms from all phased-VPol receivers coadded.
As the dominant thermal noise is independent across receivers in excellent approximation \cite{thermal-correlation}, this improves the SNR of the waveform.
All waveforms contain a single pulse with a common polarity and qualitatively similar shape, suggesting a common emission mechanism.
The bottom panel of Fig.~\ref{fig:signals} shows their power spectra and the band-defining filters \cite{PA_instrument_paper};
the signals fill the instrumented band, implying a broadband emitter.
We find qualitative agreement of the time-domain waveforms with microscopic simulations of the in-ice radio signature generated by impacting CR air 
shower cores, as Fig.~\ref{fig:signals} shows for a vertical shower induced by a $10^{17}$\,eV proton primary.
The simulation uses \corsika{} \cite{corsika8-2, corsika8} to evolve the cross-media particle cascade and \eisvogel{}~\cite{eisvogel, eisvogel_arena} for
fully electrodynamic radio emission and propagation, including transition radiation (TR) at the air-ice interface and Askaryan
radiation from the in-ice cascade.
Considering the coherently emitting charges within 1\,m of the shower axis,
we find that the majority of the shower core energy at impact is carried by high-energy photons, with a charge excess
at the surface of \mbox{$\approx$ 5\%} of the maximum in-ice charge excess.
Measurements of Askaryan radiation in the SLAC GeV-electron beam \cite{ice_askaryan} operated at a higher charge-excess fraction of $\approx$~15\%
and did not observe TR to be dominant, consistent with the quadratic scaling of radiated power with the net charge of the emitter
for coherent emission mechanisms.
Additionally, our simulations %
show that the intensity of forward-beamed TR is smaller by a factor of $>10^4$ than the intensity of on-cone Askaryan radiation,
consistent with previous studies \cite{FAERIE, coleman2024iniceaskaryanemissionair} that indicate TR \mbox{is subdominant}.

\begin{figure}[tb]
  \includegraphics[width=\columnwidth]{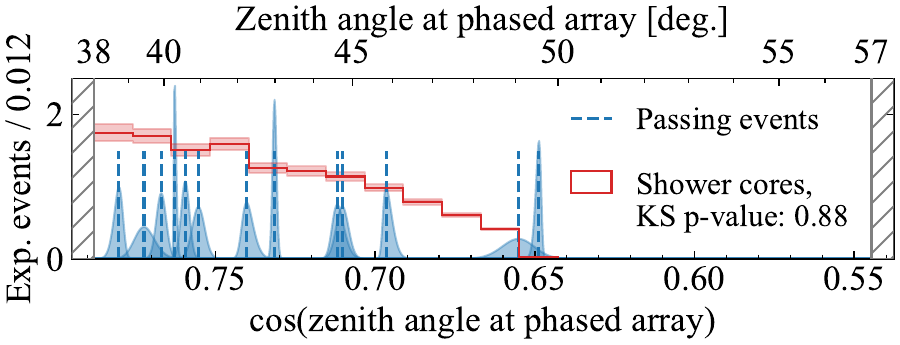}
  \caption{
	  Zenith angles measured at the phased array for the passing events (vertical lines),
	  with statistical uncertainties indicated by the blue distributions.
	  The simulated distribution for impacting shower cores, normalized to the observed yield, is shown in red.
  }
  \label{fig:zenith}
\end{figure}

\textsl{Signal arrival directions}---%
The distribution of signal arrival zenith measured at the phased array carries information about the radial location of
the source and its emission pattern.
Note that this angle is different from the CR incidence angle in the case of Askaryan radiation,
due to the $45^\circ$--$55^\circ$ (depending on depth) in-ice Cherenkov angle and refraction in the inhomogeneous ice.
Figure~\ref{fig:zenith} compares the arrival zenith angles reconstructed for the passing events with the expectation for
Askaryan radiation from impacting shower cores.
The Kolmogorov-Smirnov (KS) test yields a $p$-value of 0.88, indicating good compatibility.
Notably, no events are observed at zenith angles \mbox{$50^\circ \lesssim \theta < 57^\circ$}, where refraction in a smoothly inhomogeneous 
ice permittivity profile prevents radiation originating near the surface from reaching the detector, independently 
suggesting a near-surface in-ice source.
Here and throughout this Letter we obtain statistical uncertainties by repeatedly perturbing the observed waveforms with noise 
from minimum-bias triggers and calculating the distribution of the desired quantity.

\begin{figure}[b]
  \includegraphics[width=\columnwidth]{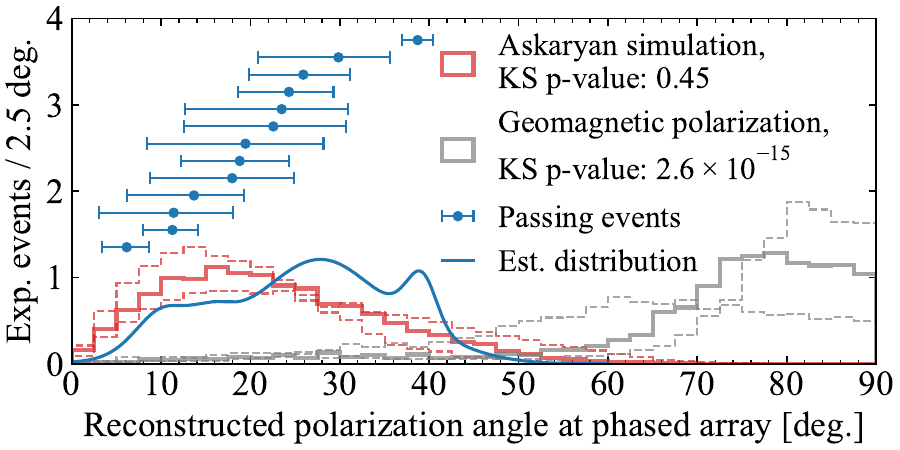}
  \caption{
    Reconstructed polarization angles for the passing events (markers, vertically offset for clarity) 
    with statistical uncertainties (error bars) and a kernel density estimate (solid blue).
    The red histogram shows the simulation prediction for in-ice Askaryan radiation from shower cores, 
    normalized to the observed event yield.
    The gray histogram shows the expectation for radiation with identical zenith angle distribution,
    but geomagnetic polarization. %
    The dashed histograms show variations of the HPol response detailed in the text.
  }
  \label{fig:polarization}
\end{figure}

\textsl{Polarization}---%
Figure~\ref{fig:polarization} shows the electric field polarization angles measured for the passing events, compared with the expectation for
in-ice Askaryan radiation from impacting shower cores and geomagnetic radiation from air showers.
This polarization measurement, distinct from the analyses in Refs.~\cite{Flaherty_pol, Salcedo_pol},
uses the HPol antenna pair at the bottom of the phased array string and its two lowermost phased-VPol
receivers, which measure the azimuthal ($E_\phi$, HPol) and polar ($E_\theta$, VPol) electric field components with respect to vertical.
To characterize the polarization, we employ the (absolute) angle $\abs{\psi} = \arctan \sqrt{\abs{f_\phi} / \abs{f_\theta}}$,
where $f_{\phi, \theta} = \int dt\, E^2_{\phi, \theta}(t) - c_{\phi, \theta}$ are the noise-subtracted fluences~\cite{aera-polarization-reconstruction},
with $c_{\phi, \theta}$ chosen to ensure that $f_{\phi, \theta}$ vanishes in the mean when evaluated on events containing no signal.
The measurement is performed in the 170\,MHz--400\,MHz band, where both antenna types have good response \cite{ARA_instrument_paper}.
Experimental systematic uncertainties on the simulated distributions are dominated by uncertainties on the in-borehole HPol antenna response.
The effect of a $\pm 50\%$ variation of the in-borehole HPol antenna effective length around the nominal response is shown by the dashed 
distributions in Fig.~\ref{fig:polarization}
\footnote{
	The cross-polarization effective length of the HPol antennas, i.e.,~their response to VPol signals, is about 10\% of their 
	copolarization effective length.
	We unfold the copolarization response in the calculation of $E^2_{\phi}$, and forward-fold the cross-polarization response in the simulation.
}.
The passing events show a mostly-$E_\theta$ signature, consistent with radial polarization and the nominal Askaryan simulation prediction.
Although we do not attempt to reconstruct the shower axis, simulation shows that the zenith region of interest includes impacting shower cores 
with inclinations between $30^\circ$ and $40^\circ$, which can generate events with measurable $E_\phi$ content.

An alternative hypothesis is in-air geomagnetic radiation, emitted by CR air showers with a characteristic 
$\mathbf{v} \times \mathbf{B}$ polarization in the plane perpendicular to the shower axis $\mathbf{v}$ and the geomagnetic field $\mathbf{B}$.
At the South Pole, the local geomagnetic dip angle is near vertical \mbox{($\approx$ ${72}^\circ$)} \cite{wmm}, leading to an $E_\phi$-dominated 
radiation signature and a polarization angle distribution different from the observed (at a KS $p$-value of $2.6\times10^{-15}$, 
cf.~Fig.~\ref{fig:polarization}).

\textsl{Test for a showerlike source}---%
Because the A5 receivers are positioned in five laterally separated strings (cf.~Fig.~\ref{fig:detector}), the view angles 
for a shallow below-surface source vary by $10^\circ$--$15^\circ$.
Across this range, the beamed
in-ice Askaryan radiation pattern is expected to produce detectable differences in the absolute radiation intensity and spectral content \cite{ZHSfreq, ice_askaryan}.
Below, we describe a procedure using signals from all receivers to identify this signature in data, probing
the spatial structure of the radiation source in a manner independent of the results presented thus far.
To illustrate the method, we consider a simple model of an in-ice cascade, in which a variable point charge $q(z)$ moving at the
speed of light along the $z$ axis represents the charge excess.
The emitted electric field amplitude at angular frequency $\om$ scales as \cite{diff_prd}
\begin{equation}
  ||\mathbf{E}(\om, \theta)|| \propto i \om \sin\theta \int dz\, q(z) \, e^{i z \frac{n \om}{c} \left(\cos\theta_c-\cos\theta\right)},
  \label{eq:emission_shower}
\end{equation}
where $\theta$ is the angle with respect to the shower axis at which the source is observed, $\theta_c$ is the Cherenkov angle,
$c$ is the speed of light in vacuum, and $n$ is the refractive index of the medium.
The Fourier integral in Eq.~\ref{eq:emission_shower} produces the signature characteristic of diffraction phenomena,
where the angular width of the radiation intensity beam pattern $\ipred{}(\om, \theta) \propto ||\mathbf{E}||^2$ is inversely proportional to the frequency.
This implies that, as $\theta$ is varied, the rate of change in the received signal intensity $\ipred{}$ scales proportionally
with frequency.
For $\om_2 > \om_1$,
\begin{equation}
  \frac{d\log \ipred{}(\om_2, \theta)}{d\theta} \biggr/ \frac{d\log \ipred{}(\om_1, \theta)}{d\theta} =
  \frac{d\log \ipred{}(\om_2)}{d\log \ipred{}(\om_1)} > 1.
  \label{eq:hflf_slope}
\end{equation}
This is in contrast to pointlike (nondiffractive) sources, such as electrically small transmitters or TR
at the air-ice interface, for which the angular radiation pattern is (approximately) independent of
frequency \cite{antenna_theory, transition_radiation1, transition_radiation2, transition_radiation3} so that
$d\log \ipred{}(\om_2) / d\log \ipred{}(\om_1) = 1$.

\begin{figure}
  \includegraphics[width=\columnwidth]{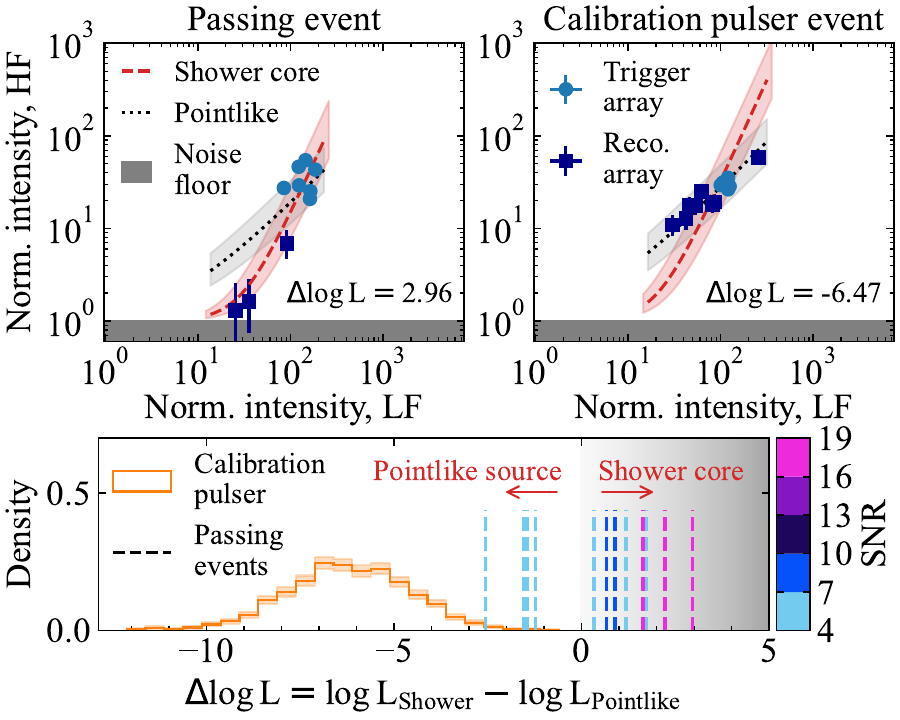}
  \caption{
    Top: noise-normalized VPol signal intensities in the LF and HF bands at multiple receivers (markers), 
    for a passing event (left)
    and a calibration pulser event (right).
    Model predictions are overlaid for impacting shower
    cores (dashed) and a pointlike source (dotted).
    Error bars show statistical uncertainties, and colored envelopes indicate the systematic effects on
    relative signal arrival direction and propagation distance described in the text.
    The respective $\Delta \log L$ values are also shown.
    Bottom: distributions of $\Delta \log L$ for the passing events (vertical lines, with average SNR at the phased array shown 
	by color) and a sample of calibration pulser events.
    }
  \label{fig:askaryan}
\end{figure}

We calculate a per-event likelihood ratio discriminant to distinguish between these two scenarios.
It is defined in terms of the (time-windowed) VPol signal intensities as measured at the digitizer in a \mbox{higher-frequency (HF)} band,
285\,MHz--400\,MHz, and a lower-frequency (LF) band, 170\,MHz--285\,MHz, normalized by the average noise intensity %
to correct for amplifier gain differences across channels.
For impacting shower cores, our simulation shows that the logarithmic signal intensities in these two bands are related in good approximation as
\mbox{$\log \ipred{}_\mathrm{HF} \sim 2.2 \log \ipred{}_\mathrm{LF}$} within about $6^\circ$ of the Cherenkov angle, while absolute radiation intensities
vary by a factor of 10 (100) in the LF (HF) band.
The discriminant is defined as $\Delta \log L = \log L_{\mathrm{Shower}} - \log L_{\mathrm{Pointlike}}$, where 
$L_{\mathrm{Shower}}$ uses the above-mentioned parametrization, and the likelihood under the pointlike hypothesis $L_{\mathrm{Pointlike}}$
assumes $\log \ipred{}_\mathrm{HF} \sim \log \ipred{}_\mathrm{LF}$ (see the End Matter). 
It is constructed symmetrically such that $\Delta \log L \approx 0$ for events not showing clear evidence for either hypothesis.
Note that differences in the signal arrival direction or the in-ice signal propagation distance between receivers are not 
corrected for but enter the likelihood as systematic uncertainties; their impact on the intensity for a shallow below-surface source
is estimated as 20\% each, much smaller than the targeted effect.
As defined, the discriminant is agnostic to the event geometry, enabling analysis of events with insufficient
information for a full reconstruction.
Its discriminating power is, however, greatest for high-SNR events that illuminate many channels.

Figure~\ref{fig:askaryan} (top left) shows the signal intensities in the two bands for %
a passing event with an average SNR at the phased array of 17.
Differences in spectral shape across A5 receivers are consistent with the expectation for strongly beamed radiation from
an impacting shower core.
This is in contrast with calibration pulser emissions (top right), which have a broader radiation pattern and 
illuminate all detector channels.
Here, the observed intensity variation is due to the angle-dependent antenna response and (to a smaller extent) 
differences in the propagation distances.
Both effects are frequency agnostic, leading to negative $\Delta \log L$ values.

The $\Delta\log L$ values of all passing events are compared in Fig.~\ref{fig:askaryan} (bottom).
Seven of the eight highest-SNR events have positive $\Delta\log L$, and the three brightest events show $\Delta\log L > 1.5$, consistently
indicating a preference for a showerlike source.
We find four low-SNR passing events with $\Delta\log L < 0$, driven by single antenna channels with %
low
LF intensity
and spectral features resembling interference fringes, cf.~Fig.~S17 in Supplemental Material~\cite{supmat-arxiv}.
These could arise from multipath propagation or reflections at the ice-air interface~\cite{c8-reflection},
which a larger event sample will help clarify.

\textsl{Conclusions}---%
With the first radio observation of CR-induced in-ice particle cascades, we here 
lay the groundwork for the analysis and simulation techniques required to study a neutrino candidate event.
Given the observed rate, the full A5 dataset should already contain more than a hundred events similar to those shown here.
Looking ahead, their Askaryan signature will \mbox{serve as} a target to develop and test detailed
electrodynamic simulation models \cite{cosmin-fdtd, arianna-surface-mode, lunar-sounders} and clarify the role of the CR background,
thus guiding the optimization of searches seeking to detect the first neutrino using a radio detector.

\textsl{Acknowledgements}---%
N.~Alden and P.~Windischhofer conducted the analysis described and were the main authors of this manuscript.
Figures were created by N.~Alden, P.~Windischhofer, and J.~Stethem. 
Text was written by N.~Alden, P.~Windischhofer, and K.~Hughes.
The events analyzed were first identified in a neutrino analysis led by K.~Hughes.
\noindent
The ARA Collaboration is grateful for support from the National Science Foundation through Award No.~2013134.
The ARA Collaboration designed, constructed, and now operates the ARA detectors. 
We would like to thank IceCube, and specifically the winterovers, for the support in operating the detector. 
Data processing and calibration, Monte Carlo simulations of the detector and of theoretical models, and data analyses were performed by a large number
of collaboration members, who also discussed and approved the scientific results presented here. 
We are thankful to Antarctic Support Contractor staff, a Leidos unit for field support and enabling our work on the harshest continent. 
We thank the National Science Foundation (NSF) Office of Polar Programs and Physics Division for funding support. 
We further thank the Taiwan National Science Council's Vanguard Program NSC 92-2628-M-002-09 and the Belgian F.R.S.-FNRS and FWO.
K.~Hughes thanks the NSF for support through the Graduate Research Fellowship Program Award No.~1746045. 
A.~Connolly thanks the NSF for Awards No.~1806923 and No.~2209588 and also acknowledges the Ohio Supercomputer Center. 
S.~A.~Wissel thanks the NSF for support through CAREER Award No.~2033500.
A.~Vieregg, C.~Deaconu, N.~Alden, and P.~Windischhofer thank the NSF for Award No.~2411662 and the Research Computing Center at the University of Chicago
for computing resources.
R.~Nichol thanks the Leverhulme Trust for their support. 
K.D.~de~Vries is supported by European Research Council under the European Union's Horizon research and innovation program (Grant Agreement 763 No.~805486). 
D.~Besson, I.~Kravchenko, and D.~Seckel thank the NSF for support through the IceCube EPSCoR Initiative (Award ID No.~2019597). 
M.S.~Muzio thanks the NSF for support through the MPS-Ascend Postdoctoral Fellowship under Award No.~2138121. 
A.~Bishop thanks the Belgian American Education Foundation for their Graduate Fellowship support.
 
\bibliography{refs}

\onecolumngrid
\section*{End Matter}
\twocolumngrid

\textsl{Background estimate details}---%
To calculate the discovery significance \cite{asymptotic_likelihood}, we use the likelihood given by
\begin{equation}
	L(\musig, \nuvec) = \prod_{r\in\{\mathrm{pass.}, \mathrm{control}\}} \Po(N^{\mathrm{obs}}_{r} | \mu_r) \times f\left(\nuvec\right),
	\label{eq:likelihood}
\end{equation}
where the product includes the region populated by the passing events and several control regions used to 
constrain backgrounds.
Each region $r$ is modeled as a Poisson counting experiment whose rate $\mu_r$ is constrained by the observed 
event count $N^{\mathrm{obs}}_r$.
For the region populated by the passing events, $\musr{} = \musig{} + \muthermal{} + \mutirbkg{} + \musurface{}$, 
where $\musig{}$ is the rate of the signal process, \muthermal{} is the rate of thermal background 
events, and \mutirbkg{} and \musurface{} are the near-horizon and on-surface background rates, 
respectively.
Systematic uncertainties are accounted for with nuisance parameters, grouped together into the vector $\nuvec{}$.
The second term in Eq.~\ref{eq:likelihood} symbolizes constraints placed on certain nuisance parameters.
The significance is calculated in terms of the profile-likelihood ratio test 
statistic $q_0 = -2 \log\left[L(\musig = 0, \doublehat{\nuvec})/L(\musigest, \hat{\nuvec})\right]$, where
`$\,\hat{}\,$' denotes the maximum-likelihood estimator (MLE) of a parameter and `$\,\doublehat{}\,$' is the conditional 
MLE, where $\musig = 0$ is fixed.

\begin{figure}[b]
  \includegraphics[width=\columnwidth]{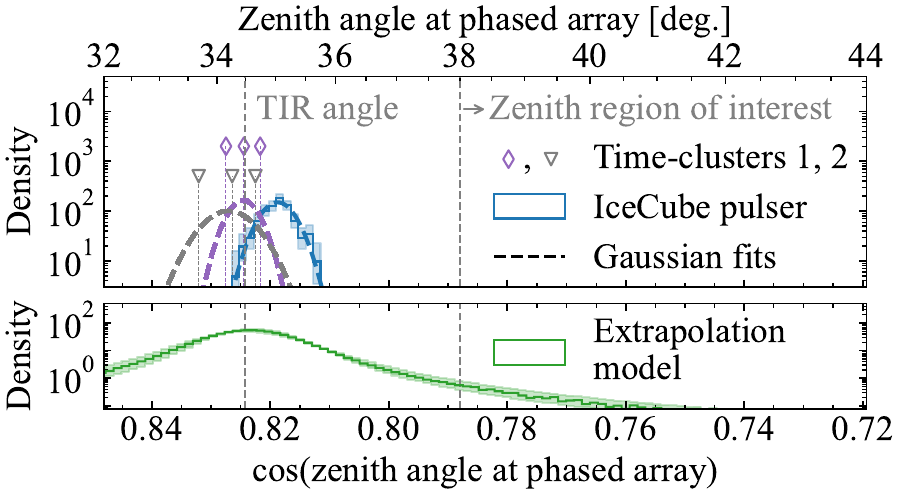}
  \caption{
    Top: zenith angles measured at the phased array for event time clusters near the TIR angle,
    with the dashed lines indicating Gaussian fits to the zenith distribution for each cluster.
    Bottom: the chosen model for extrapolating into the
    zenith region of interest (solid green) with systematic variations (light green).
  }
  \label{fig:tir_extrap}
\end{figure}

First, we address distant near-horizon sources.
Their emissions arrive at the phased array at the local TIR zenith angle, 
which may fall inside the zenith region of interest in case of significant surface roughness or unaccounted-for 
borehole tilt.
To assess this leakage, we consider the zenith distribution of time-clustered events passing the LDA selection 
in the zenith range from $32^\circ$--$38^\circ$ (``near-TIR'' control region).
Events are grouped into a cluster if they occur within a period of 8\,h, chosen to correspond to typical
anthropogenic timescales; this removes expected contributions from impacting CR shower cores.
During polar day,
this control region 
contains two clusters of three events each, shown in Fig.~\ref{fig:tir_extrap}.
Only time-unclustered events are present during polar night.
Signals from a pulsed calibration antenna on the IceCube laboratory generate a third event cluster with a mean 
zenith angle within $0.6^\circ$ of the TIR angle expected for a flat, horizontal ice surface.
We conservatively model the cluster position distribution with a (``fat-tailed'') Student's $t$ distribution 
(cf.~Fig.~\ref{fig:tir_extrap}) with parameters extracted from data.
Together with the observed cluster widths, this model results in a 
fraction $\tautir{} = \left(1.0^{+0.4}_{-0.3}\right) \times 10^{-2}$ of near-horizon 
background events entering the zenith region of interest.
In an alternative simulation-based approach making conservative assumptions on surface roughness and borehole
verticality, \tautir{} is no larger than $3 \times 10^{-3}$.
The assessed background rate in the zenith region of interest is then $\mutirbkg{} = \mutir{} \cdot \tautir{}$, where
the total event rate $\mutir{}$ in the near-TIR control region includes time-unclustered CR events,
thus overestimating the true background rate.

\begin{figure}[b]
  \includegraphics[width=\columnwidth]{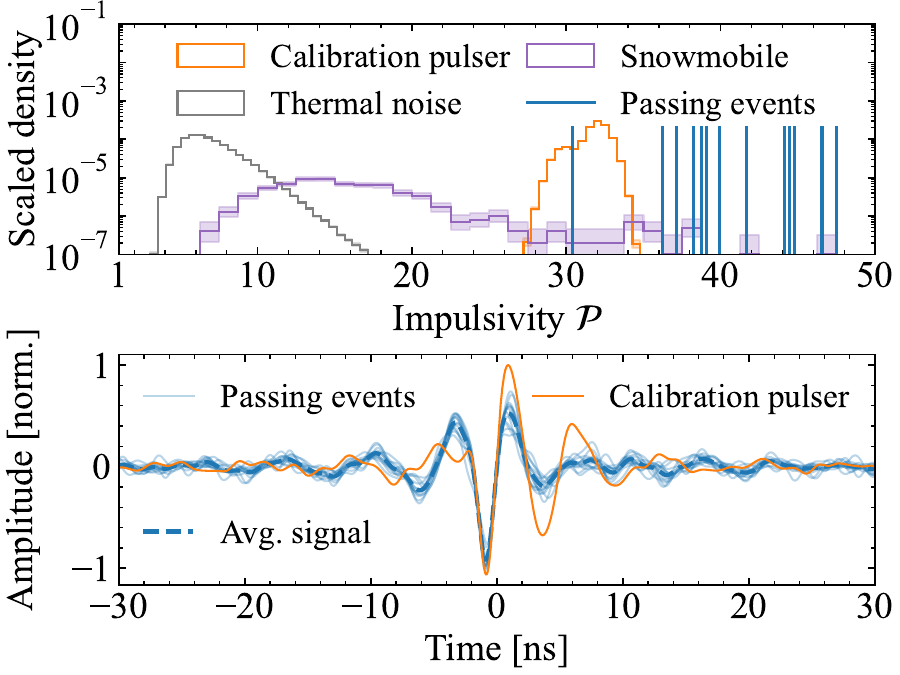}
  \caption{
    Top: impulsivity distribution for the passing events, thermal noise, and various anthropogenic sources of impulsive
    radiation, with relative normalizations adjusted for visibility.
    Bottom: comparison of VPol waveforms of the passing events and the calibration pulser.
  }
  \label{fig:event_impulsivity}
\end{figure}

Next, to estimate the rate of on-surface backgrounds, we perform an extrapolation in the zenith region of interest
in terms of signal impulsivity \imp{} and time-cluster properties.
Known anthropogenic backgrounds produce a continuous impulsivity distribution and predominantly populate the
low-impulsivity region $\imp{} < 25$, cf.~Fig.~\ref{fig:event_impulsivity}.
Signals from the calibration pulser 
show $\imp{}$ in the range 25--35, while most passing events have even higher impulsivity scores $\imp{} > 35$.
We derive a low-to-high impulsivity extrapolation factor 
$\tau_\imp{} = \mu(\imp{} > 25) / \mu(\imp{} < 25)$ for anthropogenic on-surface backgrounds 
using a control sample recorded during the passage of a GPS-equipped snowmobile near A5.
Other known sources of impulsive on-surface emission were found to lead to lower, i.e.,~less conservative,
$\tau_\imp{}$.
Similarly, a clustered-to-unclustered extrapolation factor $\tau_c = \mu(\mathrm{unclustered})/\mu(\mathrm{clustered})$ is derived in the near-TIR
control region during polar day, using the same time-cluster definition as above and emphasizing that the CR contamination present in this
region leads to an overestimation of $\tau_c$.
The estimated background rate is then $\musurface{} = \mu_{c\overline{\imp}} \, \tau_c \tau_\imp{}$,
where $\mu_{c\overline{\imp}}$ is the Poisson rate of time-clustered passing events with $\imp{} < 25$.
No events of this type are observed, leading to the upper limit on \musurface{} given in the main text.
In the significance calculation not explicitly using impulsivity information, only $\tau_c$ is applied.

\textsl{Simulation details}---%
Our Monte Carlo simulation 
uses the parametrization of the radio emission from neutrino-induced in-ice showers from
Refs.~\cite{ARZ, ARZ-param-1, ARZ-param-2}.
The model calculates the 
emitted electric field from the longitudinal charge-excess
profile of the shower and accounts for the transverse size of the cascade through form factors derived from microscopic 
simulations.
Simulated events are weighted by the CR flux at the ice surface, parametrized in terms of the energy 
deposited in the ice.
This flux is calculated using the CR flux and energy evolution of $\mathrm{X_{max}}$ measured by the Pierre Auger 
Observatory \cite{Auger_xmax_vs_energy}, as derived in Refs.~\cite{coleman2024iniceaskaryanemissionair, alan-personal}.
Comparing the signals produced by 
this
approach to microscopic simulations of impacting CR shower 
cores \cite{coleman2024iniceaskaryanemissionair}, we find that the
relative signal amplitude scale factor depends strongly on the view angle at which the in-ice cascade
triggers.
In the region around $3^\circ$ off the Cherenkov angle, which contributes about 80\% of the passing events,
it varies from 0.4--0.8, resulting in the rate interval of 8--34\,yr$^{-1}$ (cf. Fig. S15 in Supplemental Material~\cite{supmat-arxiv}).
If the flux measured by the Telescope Array \cite{ta-flux} is used instead, the rate prediction increases by $\approx$10\%.

\begin{figure}[b!]
  \includegraphics[width=\columnwidth]{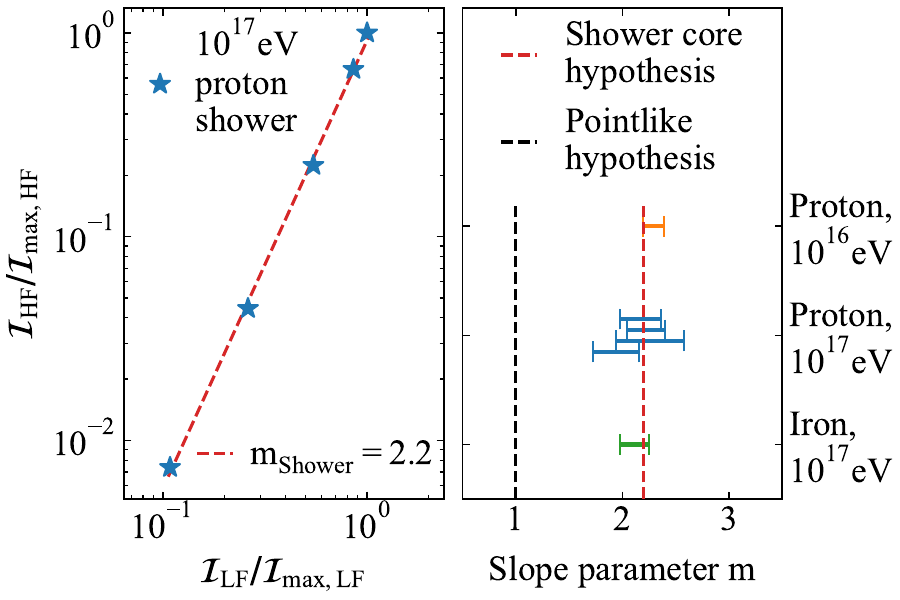}
  \caption{
    Left: simulated relative signal intensity in the LF and HF frequency bands for multiple antennas (shown by stars) with view angles within $\approx6^\circ$ around the approximate in-ice Cherenkov angle,
        shown for a vertical $10^{17}\,\mathrm{eV}$ proton shower.
    Right: range of the slope parameter $m$ for vertical showers with different energies, random seeds, and primaries.
	In each case, the range is calculated by finding the minimum and maximum slope parameter between any pair of antennas.
  }
  \label{fig:hflf_simulation}
\end{figure}

Microscopic simulations of ice-impacting showers are performed with \corsika{} and \eisvogel{}, using
a five-layer atmospheric density model and an ice density profile
consistent with the refractive
index profile $n(z)$ derived in Ref.~\cite{ara-lowthresh-analysis}.
The signal observed at the phased array is calculated by convolving the electron and positron tracks in the cascade
with an electrodynamic Green's function \cite{eisvogel}, which encapsulates signal emission, propagation, and reception.
Figure~\ref{fig:hflf_simulation} shows the simulated signal intensity received at different view angles
in the LF and HF frequency bands.
The scaling $\log \ipred_{\mathrm{HF}} \sim m \log \ipred_{\mathrm{LF}}$ between the logarithmic intensities is apparent 
with a slope parameter $m$ in the range 2--2.5 across showers of different primary energy and mass composition.

\textsl{Test for a showerlike source}---%
To probe the nature of the emission mechanism, we use a model of the form
\begin{equation}
  \ipredlf{}(s; m, b) = \exp(s),\,\,\,\, \ipredhf{}(s; m, b) = \exp(m\, s + b),
  \label{eq:hflf_model}
\end{equation}
where $s\in\mathbb{R}$ is a curve parameter and $b$ controls the relative normalization between the two bands.
From Fig.~\ref{fig:hflf_simulation}, we find a mean value of $m_{\mathrm{Shower}} = 2.2$ across different 
shower realizations.
We use $m_{\mathrm{Pointlike}} = 1.0$ to describe a pointlike source.
In terms of the observed noise-normalized intensities ${I}_{B}^{(c)}$ in channel $c$ and
band $B\in\{\mathrm{LF}, \mathrm{HF}\}$, the likelihood takes the schematic form
\begin{equation}
  L(m) = 
  \sup_{\alphavec} 
  \prod_{c, B, D} \,\,
  \gamma\left({I}_{\mathrm{B}}^{(c)}\bigg|1 + 
  \frac{\ipred_{B}(s_c; m, b_D)}{\ipred{}_{B,\mathrm{noise}}} \, \mathcal{S}_B(\Delta b_c)\right),
	\label{eq:hflf_likelihood}
\end{equation}
where the nuisance parameters $\alphavec = \{\{s_c\}, \{b_D\}, \{\Delta b_c\}\}$ are profiled.
In Eq.~\ref{eq:hflf_likelihood}, $\ipred{}_{B,\mathrm{noise}}$ is the mean noise intensity in band $B$ and
$\gamma({I}|\ipredn{})$ is the probability of observing a noise-normalized intensity ${I}$ given a
predicted (mean) normalized intensity $\ipredn{}$.
It represents the statistical uncertainty and is approximated by an independent gamma distribution for each channel,
obtained using the procedure described in the main body.
Introducing a nuisance curve parameter $s_c$ for each channel naturally accounts for statistical
uncertainties in both bands.
Additional systematic effects are accounted for by per-channel normalization parameters $\Delta b_c$, constrained by a Gaussian term
in the likelihood.
They implement per-band multiplicative corrections $\mathcal{S}_B$ of $\pm 40\%$ at $1\,\sigma$ in the direction orthogonal to the 
parametric curve in Eq.~\ref{eq:hflf_model}.
This is sufficient to cover the effects listed in the main body.
Because channels are read out by one of two DAQ systems $D$ with different architectures,
we use independent normalization
parameters $b_D$ for each digitizer, i.e.~compare only relative signal intensities measured with the same system.
Only channels with signal-to-noise power ratio $> 4.5$ in the HF $\cup$ LF band
and not saturating the dynamic range of the digitizer are included in Eq.~\ref{eq:hflf_likelihood} for a given event.

\mbox{}
\clearpage

\onecolumngrid
\begin{center}
{
\large\bfseries
Supplemental Material for\\``Observation of In-ice Askaryan Radiation from High-Energy Cosmic Rays''\\[4mm]
}
The ARA Collaboration
\end{center}
\vspace{5mm}
\twocolumngrid

\setcounter{equation}{0}
\setcounter{figure}{0}
\renewcommand{\thefigure}{S\arabic{figure}}
\renewcommand{\theequation}{S\arabic{equation}}

\section{Event reconstruction}
\noindent
The primary event reconstruction is performed using the phased-array antenna channels.
We create an interferometric reconstruction map $C(R, z)$ as defined in Eq.~3 of Ref.~\cite{ara-lowthresh-analysis}.
It contains the cross-correlation, averaged over all phased-array antenna pairs, calculated under the hypothesis of
a point source at radial coordinate $R$ (with origin at the phased-array borehole) and vertical coordinate $z$
(with origin at the ice surface).
The values are normalized to be in the interval $[-1, 1]$, with larger positive values indicating better compatibility
of the observed waveforms with the source hypothesis.
The construction of $C(R, z)$ requires knowledge of the signal propagation times from the hypothesized source location to
the antennas, which we compute in geometric optics by solving the eikonal equation with the 
fast marching method \cite{fast_marching_method, Pykonal} on the refractive-index model in 
Eq.~1 of Ref.~\cite{ara-lowthresh-analysis}.
For near-surface in-ice sources, two (nearly-degenerate) ray solutions exist.
We only consider the ray with the lower propagation time; the solution undergoing a reflection at the ice-air interface
is not considered.

\begin{figure}[b]
  \includegraphics[width=0.8\columnwidth]{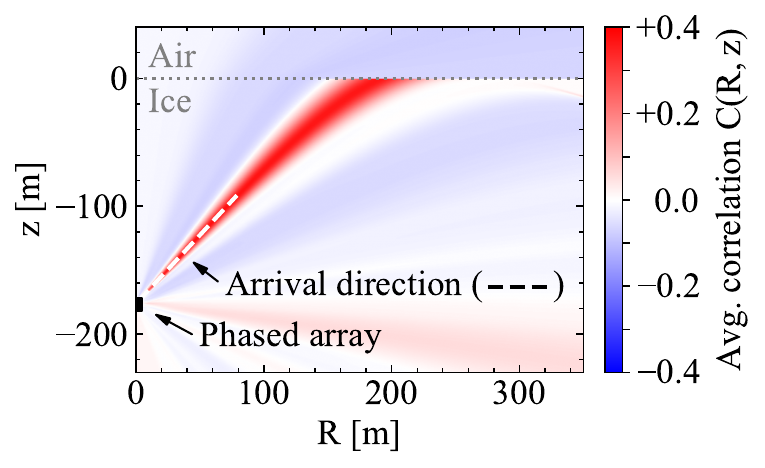}
  \caption{
	  Example reconstruction map $C(R, z)$ for a passing event, with the location of the phased array
	  and the signal arrival direction indicated on the plot.
  }
  \label{fig:recoplot}
\end{figure}

An example reconstruction map is shown in Fig.~\ref{fig:recoplot}.
The set of points $\{(r, \mathrm{argmax_z} C(r, z))\}$ represents the family of point source locations 
which best match the observed data, and its slope $dz/dr$ extrapolated to $r = 0$ corresponds to the radiation arrival 
direction at the phased array location.
Using this method, we find typical statistical uncertainties of $0.05^\circ$--$0.5^\circ$ depending on SNR, 
and estimate an overall zenith accuracy of $\mathcal{O}(1^\circ)$, obtained by reconstructing a far-field calibration 
source (cf.~Fig.~6).
An alternative reconstruction which assumes planar wavefronts finds radiation zenith angles consistent with the above 
within the stated precision.

We additionally create a reconstruction map $C(x, y)$ in the horizontal plane at $z = -5$\,m to reconstruct the azimuthal
source location.
This map makes use of the signals detected by the antennas in the four outer strings, collectively referred to as
the reconstruction array.

\section{Event properties}
\begin{figure}[tp]
	\includegraphics[width=\columnwidth]{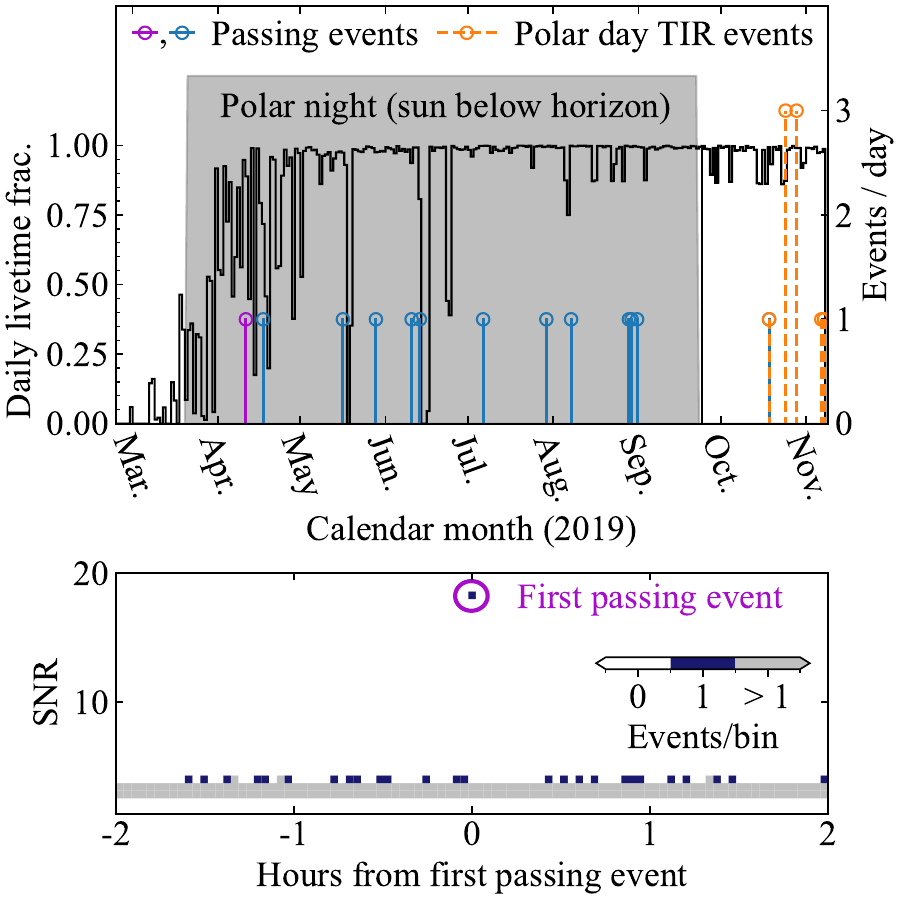}
	\caption{
		Top: Daily detector livetime fraction for the analyzed dataset (left axis).
		The gray region indicates polar night (March~21--Sept.~21, the period during which the sun remains below the horizon).
		Vertical solid lines indicate the passing events in the zenith region of interest 
		(with zenith arrival angle in the region $38^\circ$--$57^\circ$).
		Orange vertical dashed lines show events in the near-TIR control region 
		($32^\circ$--$38^\circ$) during polar day.
		Bottom: The per-antenna SNR averaged over the phased-VPol receivers for RF-triggered events in a 
		four-hour period centered on the first passing event, also indicated by purple in the top panel.
	}
	\label{fig:event_timing}
\end{figure}
\noindent
\paragraph{Event arrival times}
Fig.~\ref{fig:event_timing} shows the event arrival times for the 13 passing events along with the data-taking
status of the phased-array trigger (top panel).
All events occur on different calendar days and surrounding triggers are consistent with thermal noise (bottom panel).
Also shown is the near-TIR control region during polar day,
which contains six time-clustered and three time-unclustered events and is used 
in the construction of the near-horizon background estimate.

\begin{figure}[tp]
	\includegraphics[width=\columnwidth]{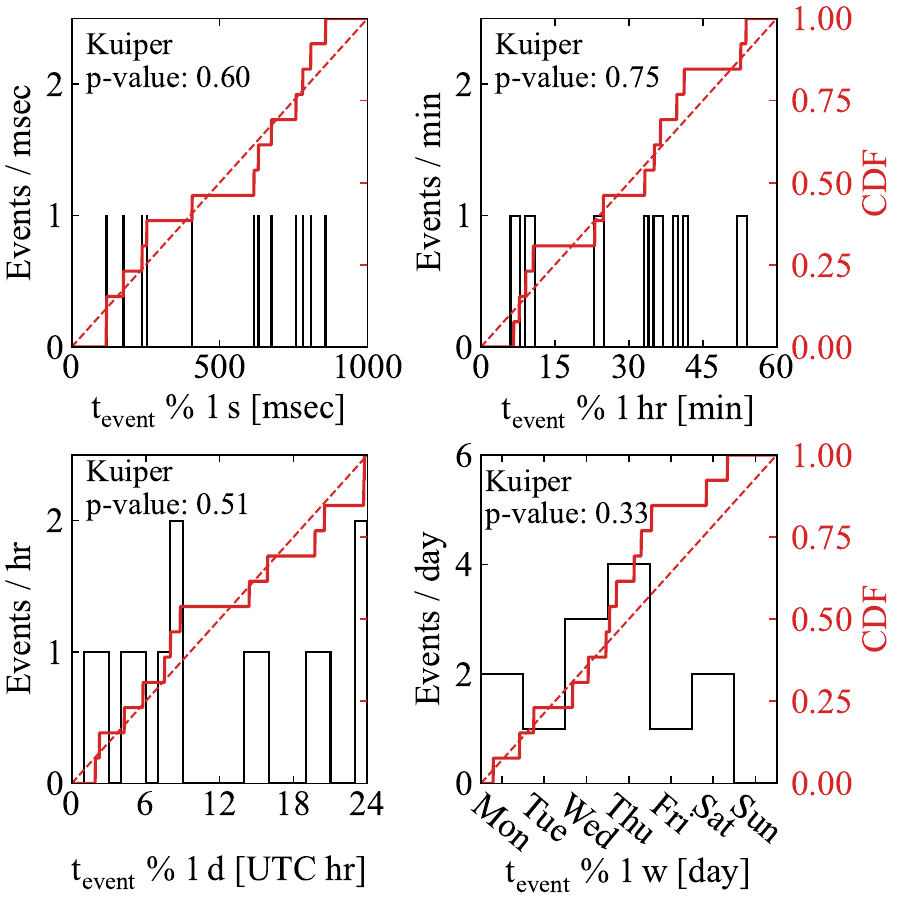}
	\caption{
		Kuiper's test for periodicity on different anthropogenic time scales: UTC second (top left), 
		UTC hour (top right), day (bottom left), and calendar week (bottom right).
		Black lines show the binned event rate and solid red lines indicate the CDFs of the event arrival times.
		The Kuiper $p$-value is calculated for the null hypothesis of a uniform distribution, 
		whose CDF is shown by the dashed red lines.
	}
	\label{fig:periodicity}
\end{figure}

\paragraph{Test for periodicity}
We use Kuiper's test to test for possible periodicities in the event arrival times,
which could be indicative of an artificial origin of the passing events.
We perform the test for periods of calendar week, day, UTC hour, and second, cf.~Fig.~\ref{fig:periodicity}.
In all cases, we find the data to be consistent with aperiodicity.

\begin{figure}
	\includegraphics[width=\columnwidth]{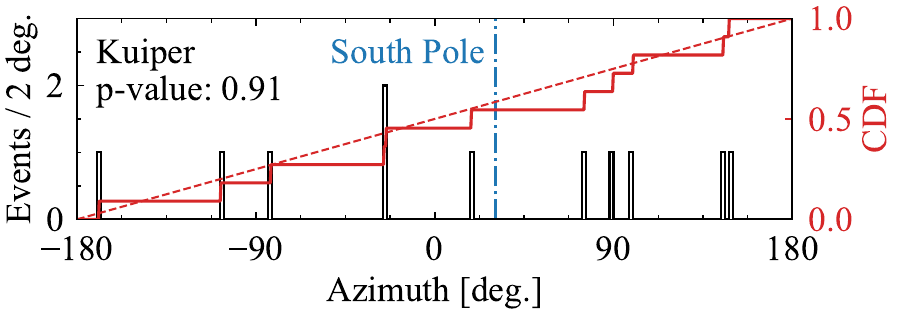}
	\caption{
		Binned reconstructed azimuth for eleven of the thirteen passing events (black),
		measured counterclockwise.
		For two events, there is insufficient information in the reconstruction array channels to find a unique azimuth solution.
		Solid red lines indicate the data CDF, and the Kuiper $p$-value is calculated for the null hypothesis of
		a uniform distribution, whose CDF is shown by the dashed red line.
		The direction to South Pole is indicated by the blue dash-dotted line.
	}
	\label{fig:azimuth}
\end{figure}

\paragraph{Azimuthal event distribution}
Fig.~\ref{fig:azimuth} shows Kuiper's test for the reconstructed azimuthal event directions. 
The test returns a $p$-value of 0.91, indicating consistency with a uniform distribution in azimuth.

\begin{figure}[tp]
	\includegraphics[width=\columnwidth]{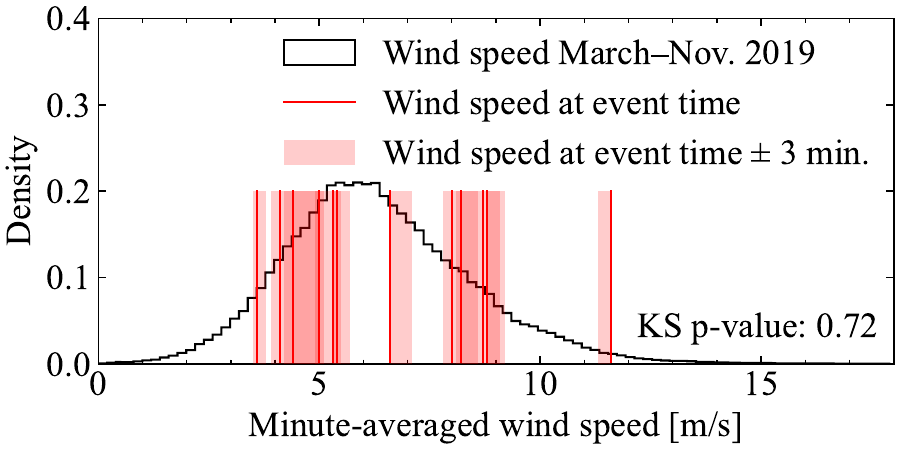}
	\caption{
		Minute-averaged wind speeds measured at South Pole Station \cite{wind-data} at the times
		of the passing events 
		(vertical red lines) and during a window of 6 min centered on the event time 
		(shaded vertical regions).
		The black histogram shows the inclusive distribution of minute-averaged wind speeds
		from April--November 2019.
		The KS $p$-value is calculated for the null hypothesis of both distributions being identical.
	}
	\label{fig:wind_speed}
\end{figure}

\paragraph{Wind speed distribution}
Surface charges known to be produced by the triboelectric effect during periods of high winds \cite{triboelectric} 
can generate time-clustered impulsive emissions at wind speeds $\gtrsim$ 10\,m/s, with
some location-dependence.
Fig.~\ref{fig:wind_speed} compares the wind speeds measured at the times of the passing events with the 
inclusive distribution for the period of the analyzed dataset.
The two-sample Kolmogorov-Smirnov (KS) test between the two distributions returns a $p$-value of 0.72, 
consistent with a wind-speed-independent mechanism.

\section{Background estimate}
\begin{figure}[tp]
\centering
\includegraphics[width=\columnwidth]{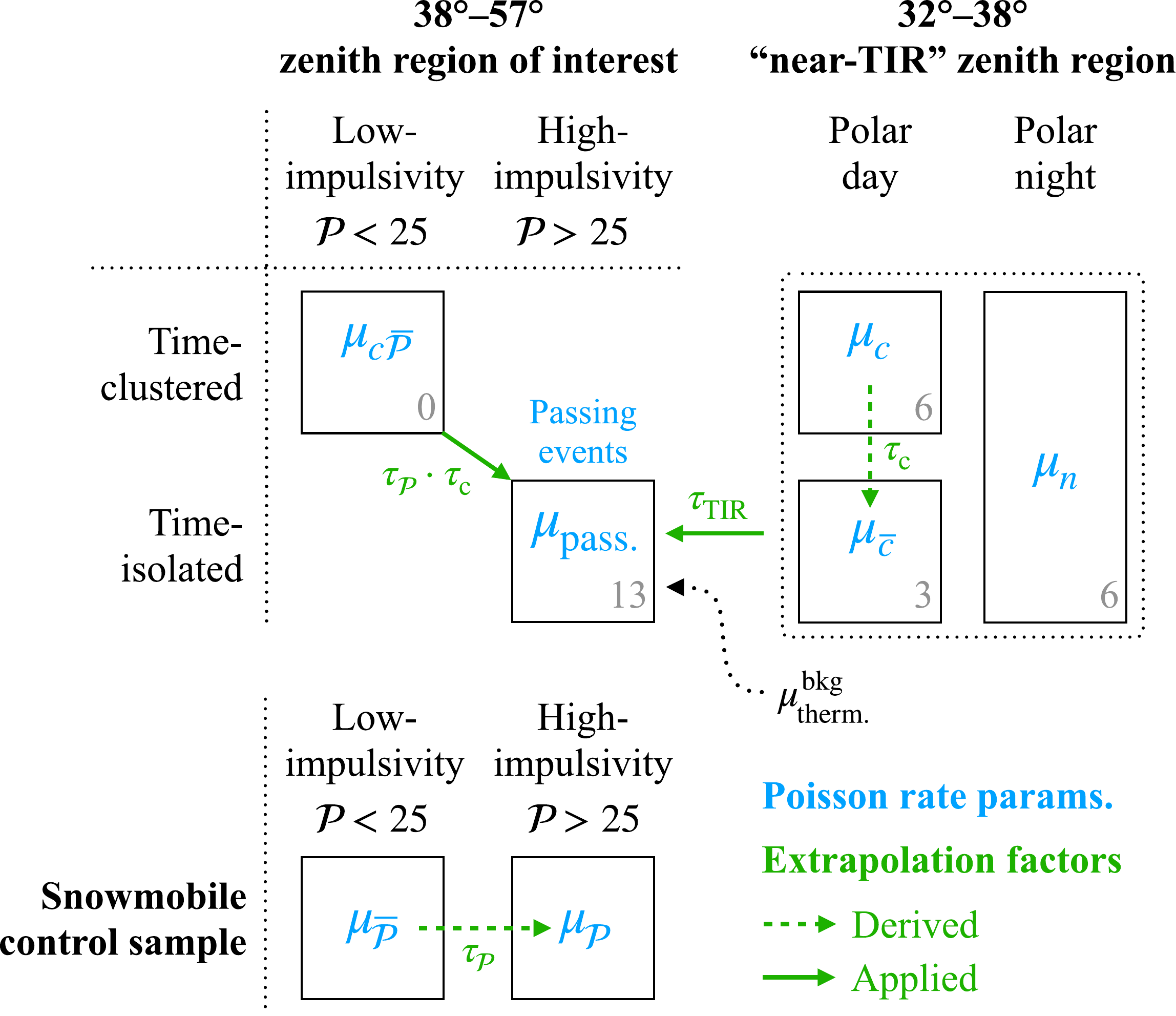}
\caption{
	Graphical depiction of the combined likelihood used in the significance calculation.
	The likelihood includes event regions defined in terms of zenith angle, time-clustering properties, 
	and impulsivity $\imp{}$.
	Each region $r$ is modeled in the likelihood as a Poisson counting experiment with rate 
	parameter $\mu_r$, while extrapolation factors $\tau$ relate the rate parameters of different regions.
	The LDA selection cut is applied in all regions shown.
	The assessed background rate of thermal-noise events, $\muthermalbkg{}$, is extracted from the sample of events
	failing the LDA cut and enters the likelihood as a constrained nuisance parameter.
	The numerical values represent the observed event yields in each region.
}
\label{fig:likelihood_structure}
\end{figure}

\noindent
To assess the contributions from thermal noise, distant near-horizon sources, and impulsive 
on-surface sources, we use multiple control regions enriched in these respective backgrounds.
Information from all regions is incorporated into a global likelihood to calculate 
the discovery significance \cite{asymptotic_likelihood}.
Fig.~\ref{fig:likelihood_structure} gives a graphical summary of all regions entering the likelihood, with
further details on the background estimate and statistical inference procedures provided in the following.

\subsection{Thermal noise}
\begin{figure}
	\includegraphics[width=\columnwidth]{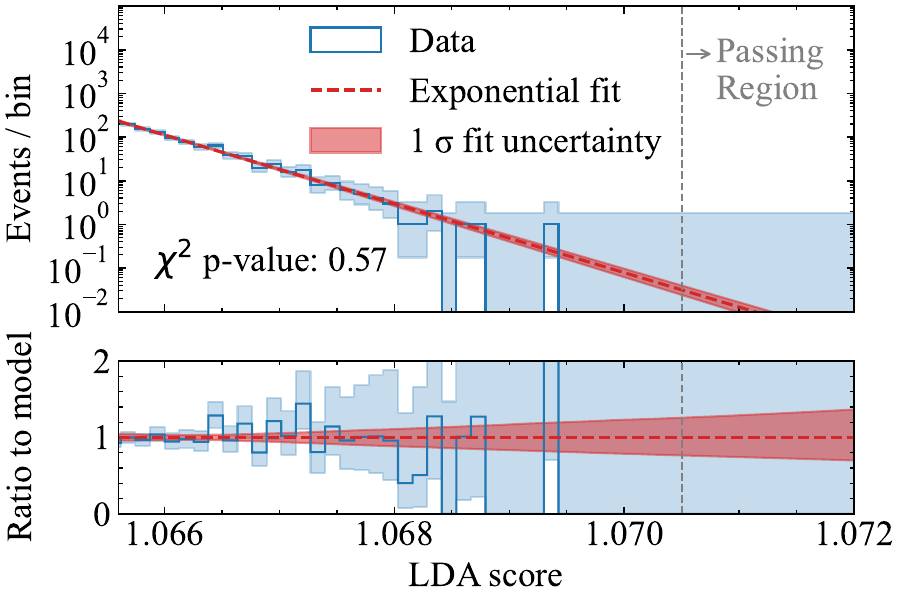}
	\caption{
		Distribution of the linear discriminant (LDA) score for events failing the LDA cut in the zenith
		region of interest, with the fitted exponential model overlaid (dashed red) and the red band
		showing the 68\% uncertainty envelope.
		The goodness-of-fit is measured by the $p$-value calculated for the likelihood-ratio test statistic
		according to Ref.~\cite{cousins-gof}.
	}
	\label{fig:bkg_thermal}
\end{figure}

\noindent
To estimate the background from thermal-noise events, we closely follow the procedure outlined in 
Ref.~\cite{ara-lowthresh-analysis}.
We perform a binned maximum-likelihood fit of an exponential model to the population of events failing the LDA cut
in the zenith region of interest, cf.~Fig.~\ref{fig:bkg_thermal}.
The binning is chosen using the Freedman-Diaconis rule.
The estimated background rate $\muthermalbkg$ is obtained as the integral of the fitted yield model in the 
passing region.
Propagating the postfit uncertainties on the fit parameters as described in Ref.~\cite{ara-lowthresh-analysis}
leads to the estimate $\muthermalbkg = 0.14^{+0.05}_{-0.03}$, which enters the likelihood as a nuisance parameter
with a log-normal constraint term.

\subsection{Distant near-horizon sources}
\noindent
To characterize distant near-horizon background sources, we use time-clustered events which reconstruct near the 
TIR angle expected for a flat ice surface (``near-TIR control region'').
We then construct an extrapolation model to assess their leakage into the zenith region of interest.

\subsubsection{Derivation of the extrapolation factor \tautir{}}
\noindent
Each event cluster observed in the near-TIR control region during polar day is characterized by the mean zenith 
position, $\cos\thetaclus$, and the standard deviation of its events in $\cos\theta$.
We choose a generalized Student's $t$ distribution to model the unknown process generating the cluster positions
$\cos\thetaclus$, defined as
\begin{equation}
	\cos\thetaclus \sim \avg{\cos\thetaclus} + s \sqrt{1+\frac{1}{n}} \, t_{n-1},
	\label{eq:pdf_cluster_mean}
\end{equation}
where $n$ is the number of observed clusters.
The parameter $\avg{\cos\thetaclus} := \sum_{i=1}^n \cos\theta_{\mathrm{clust.},i} / n$ is the sample mean over all observed cluster 
positions, $s^2 = \sum_{i=1}^n \left(\cos\theta_{\mathrm{clust.},i} - \avg{\cos\theta}\right)^2 / (n-1)$ is the sample variance,
and $t_{n-1}$ is a random variable distributed according to a Student's $t$ distribution with $n-1$ degrees of 
freedom.
The model in Eq.~\ref{eq:pdf_cluster_mean} is constructed to reduce to a normal distribution in the limit of many 
cluster observations, but exhibits long tails for finite $n$.
It therefore conservatively assigns a higher extrapolation factor in our case, 
where there are only three observed clusters ($n = 3$, see Fig.~6) and the true cluster distribution family cannot be inferred from data.
Below, we compare this choice to a modeling-based approach to demonstrate that it yields conservative results for the
extrapolation factor.
We also note in passing that a distribution of a form similar to Eq.~\ref{eq:pdf_cluster_mean} arises in the expression 
for the prediction interval of a random variable distributed according to a normal distribution with unknown mean and 
variance \cite{predictive-inference}.

The final extrapolation model is given by the distribution of cluster positions
in Eq.~\ref{eq:pdf_cluster_mean}, convolved with a normal distribution with a standard deviation corresponding to
the maximum observed cluster standard deviation.
The zenith extrapolation factor $\tautir{}$ is then calculated as 
$\tautir{} = p(\mathrm{zenith\,\,ROI}) / p($near-TIR$)$, where $p(\mathrm{zenith\,\,ROI})$ and $p($near-TIR$)$ 
are the probabilities assigned by this model to the zenith region of interest and the near-TIR control region, 
respectively.
Systematic uncertainties on $\tautir{}$ arising from the zenith angle reconstruction are assessed
by varying the measured zenith angles within the observed cluster width.
For clusters that are produced by stationary (or slowly-moving) sources, the cluster width estimates the detector
resolution for the clustered events.
The resulting distribution for $\tautir{}$ is approximated by a gamma distribution 
which is
used as a constraint term for
the nuisance parameter $\tautir{}$ in the likelihood.

\paragraph{Comparison to modeling}
Long tails in the event-cluster zenith distribution can be generated, for example, if the ice surface exhibits
significant roughness or the borehole housing the phased-array antennas deviates from vertical.
We consider both cases in turn.
For an in-air source, the radiation refracted into the ice probes the surface on a length scale given by the
interface Fresnel zone (IFZ) \cite{fresnel}.
In the relevant case of grazing incidence on the surface, the extent of the IFZ along the radiation propagation 
direction becomes of the order of the horizontal separation between the source and the receiver.
This implies that only surface features on scales of hundreds of meters are relevant.
To estimate the surface roughness at these scales, we use data from a GPS receiver mounted on a snowmobile
traversing the radial connecting South Pole Station and A5.
Using the recorded elevation relative to the WGS 84 reference ellipsoid, we estimate a surface roughness of 
$\approx$ $0.5^\circ$ rms at a length scale of 100\,m.
This value is consistent with 
measurements taken with an inclinometer near South Pole Station at the 
1\,m length scale and then averaged~\cite{inclinometer}.

\begin{figure}[tp]
\centering
	\includegraphics[width=\columnwidth]{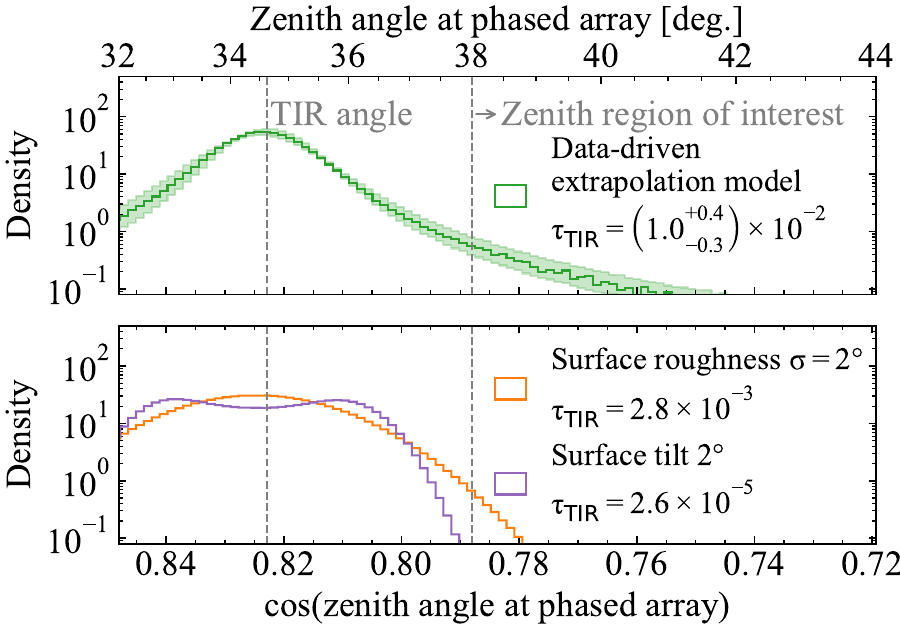}
	\caption{
		Top: The chosen data-driven zenith extrapolation model for the calculation of $\tautir{}$ (solid green) with systematic variations (light green), reproduced from Fig.~6.
		Bottom: Zenith distribution expected for events from near-horizon sources under different assumptions 
		on the roughness and tilt of the surface or, equivalently, the borehole.
	}
	\label{fig:tir-comparison}
\end{figure}

Fig.~\ref{fig:tir-comparison} shows the results of a simple Monte Carlo simulation, in which we place near-horizon
sources uniformly at all azimuths and randomly vary the local orientation of the ice surface.
Even assuming a surface roughness of $2^\circ$ (rms), the expected extrapolation factor is $\tautir \approx 3\times 10^{-3}$,
a factor of three smaller than the data-driven value used in the main body.
A similar result holds if the borehole deviates from the vertical, or, equivalently, the surface exhibits
an overall slope.

\subsubsection{Application of the extrapolation factor \tautir{}}
\noindent
To find an upper bound on the event yield in the zenith region of interest contributed from near-horizon event sources, 
we normalize the above extrapolation model by the total yield observed in the near-TIR zenith range.
We explicitly include time-isolated events, for which a contamination from impacting CR shower cores is expected.

With these assumptions, the background rate extrapolated into the zenith region of interest is 
then $\mutirbkg{} = \tautir{} \, \mutir{}$, where $\mutir{}$ is the total rate parameter in the near-TIR 
control region (cf.~Fig.~\ref{fig:likelihood_structure}).
It is given as the sum of the rates of the time-clustered ($\muclust{}$) and time-isolated ($\muunclust{}$) events during 
polar day, as well as the total event rate during polar night ($\munight{}$),
$\mutir{} = \muclust{} + \muunclust{} + \munight{}$.

\subsection{Impulsive on-surface sources}
\noindent
Nearby on-surface background sources produce events that evanescently couple into the ice \cite{Jackson_EM} and 
may therefore reconstruct into the zenith region of interest.
Depending on their location on and possible movement across the surface, such sources may show a variety of zenith angle 
distributions, making a geometric extrapolation similar to the above difficult.
Instead, we exploit the fact that (anthropogenic) background sources typically produce emissions that cluster in time
and, as shown below, exhibit a very broad impulsivity distribution.
To assess the contribution of such background events similar to the passing events, we therefore
perform an extrapolation in terms of time-cluster size and impulsivity.
Impulsivity is quantified by the ratio of maximum instantaneous signal power to mean signal power, cf.~Ref.~[27].

\subsubsection{Derivation of the extrapolation factors \tauclust{} and \tauimp{}}
\noindent
To extract the time-cluster extrapolation factor $\tauclust{}$, we consider events in the near-TIR control region 
during polar day.
In this period, the sun is above the horizon and activity levels at nearby South Pole Station are elevated, thereby enriching
this control region in typical anthropogenic signal activity.
Utilizing the same cluster definition as in the derivation of the near-TIR background estimate, we observe
six clustered and three unclustered events (cf.~Fig.~\ref{fig:event_timing}).
Assuming that all these events are generated by background processes, we have 
$\tauclust{} = \muunclust{} / \muclust{} = 0.50^{+0.49}_{-0.26}$, where the rate 
parameters $\muunclust{}$ and $\muclust{}$ 
refer to time-unclustered and time-clustered events, respectively.
We note that this is a conservative assumption and likely to overestimate the true value of the 
clustered-to-unclustered extrapolation factor due to contributions from time-unclustered CR events, which are included
in $\muunclust{}$.

\begin{figure}
	\includegraphics[width=\columnwidth]{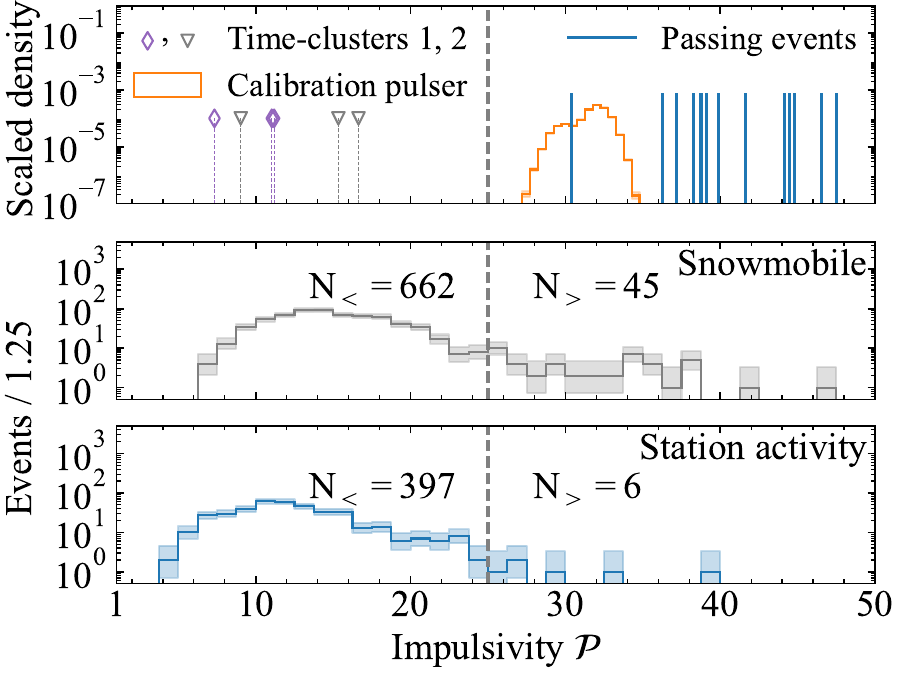}
	\caption{
		Top: Impulsivity scores of time-clustered events in the near-TIR control region (cf.~Fig.~6),
		signals from the calibration pulser, and the passing events.
		Center: Impulsivity score distribution of events generated by a snowmobile maneuvering near A5.
		Bottom: Impulsivity score distribution of time-clustered surface events generated during known
		station activity.
		$N_<$ ($N_>$) refers to the number of events with impulsivity scores less (greater) than $\imp{} = 25$.
	}
	\label{fig:impulsivity}
\end{figure}

Similarly, we define an impulsivity extrapolation factor $\tauimp{}$ by considering two event samples enriched in 
typical (known) on-surface anthropogenic activity, both shown in Fig.~\ref{fig:impulsivity}.
This includes a sample of triggers recorded in a dedicated calibration run during which a GPS-equipped snowmobile
was maneuvering close to A5, and time-clustered triggers produced during known station activity including remote
configuration changes.
In both cases, we observe a broad impulsivity distribution in which more than 90\% of all events show low impulsivity
scores (defined by $\imp{} < 25$), but a long tail exists that extends to high impulsivities ($\imp{} > 25$)
similar to those of the passing events.
The snowmobile dataset exhibits a higher ratio $N_> / N_<$ of high-impulsivity to low-impulsivity events, so we
conservatively use this event sample to define the extrapolation factor as 
$\tauimp{} = \mu_{\imp} / \mu_{\overline{\imp}} = \left(6.8^{+1.1}_{-1.0}\right) \times 10^{-2}$, 
where $\mu_{\imp}$ ($\mu_{\overline{\imp}}$) is the Poisson rate parameter for high-impulsivity (low-impulsivity) 
snowmobile events.
The time-clustered events in the near-TIR control region display low impulsivities consistent with the above control samples.

\subsubsection{Application of the extrapolation factors \tauclust{} and \tauimp{}}
\noindent
Under the assumption that the clustered-to-unclustered event rate for backgrounds is independent
of the zenith angle, we may apply $\tauclust{}$ (derived in the near-TIR zenith range) in the zenith region of interest.
If we further assume that the impulsivity of a background event is independent of whether or not it exists in a cluster, 
we may multiplicatively apply $\tauclust{}$ and $\tauimp{}$ to find
the estimated background rate in the region populated by the passing events as 
$\musurfacebkg{} = \tauclust{}\,\tauimp{}\,\muclustlowimp{}$.
In this expression, $\muclustlowimp{}$ is the rate parameter for low-impulsivity, time-clustered passing events
observed in the zenith region of interest.
There are no observed events of this kind, resulting in a Feldman-Cousins
upper limit on $\musurfacebkg{}$ of 0.12 at 95\% confidence level,
also given in the main text.

This upper limit is calculated by performing the Neyman construction using the profile-likelihood ratio ordering
principle~\cite{pdg-statistics, fc_confidence} 
\begin{equation}
	q_{\musurfacebkg{}} = -2 \log \frac{\Lsurf(\musurfacebkg{}, \doublehat{\nuvec}(\musurfacebkg{}))}{\Lsurf(\hat{\mu}_{\mathrm{surf.}}^{\mathrm{bkg}}, \hat{\nuvec})},
\end{equation}
where the vector of nuisance parameters $\nuvec = \{\mulowimp{},\,\muhighimp{},\,\muunclust{},\,\muclust{}\}$ contains
the independent Poisson rates of the participating event regions 
$r \in \{\lowimp{},\,\highimp{},\,\clustlowimp{},\,\clust{},\,\isolated{}\}$.
The notation `$\,\hat{}\,$' represents the maximum-likelihood estimator (MLE) of a parameter
and `$\,\doublehat{}\,$' is the conditional MLE.
The expression $\doublehat{\nuvec}(\musurfacebkg{})$ in the numerator is the value of \nuvec{} that maximizes \Lsurf{}
for a given fixed value of $\musurfacebkg{}$. 
The likelihood $\Lsurf$ models each region as a Poisson counting experiment,
\begin{equation}
	\Lsurf\left(\musurfacebkg{}, \nuvec\right) = \prod_{r} \Po(N^{\mathrm{obs}}_{r} | \mu_r).
	\label{eq:surface_likelihood}
\end{equation}

\subsection{Significance calculation}
\noindent
With the above estimates for the background, the rate parameter for the region populated by the passing events,
i.e.~time-isolated, high-impulsivity events in the zenith region of interest, is
$\musr = \musig + \muthermalbkg + \mutirbkg + \musurfacebkg$, where $\musig$ measures the contribution 
from the targeted physics signal process.
The full expression for the likelihood $L(\musig, \nuvec)$ is
\begin{equation}
	L(\musig, \nuvec) = \left[\prod_{r} \Po(N^{\mathrm{obs}}_{r} | \mu_r)\right] \times f\left(\muthermalbkg, \tautir\right),
	\label{eq:full_likelihood}
\end{equation}
where the product now includes all event regions $r \in \{\mathrm{pass.},\,\lowimp{},\,\highimp{},\,\clustlowimp{},\,\clust{},\,\isolated{},\,\nosun{}\}$
and the second term represents the external constraints 
\begin{equation}
	f\left(\muthermalbkg{}, \tautir{}\right) = \mathrm{Lognorm}\left(\muthermalbkg{}\right) \times \mathrm{Gamma}\left(\tautir{}\right).
\end{equation}
In the following, the signal rate $\musig$ is referred to as the parameter of interest, 
while the nuisance parameters are $\nuvec = \{\muthermalbkg{},\,\tautir{},\,\mulowimp{},\,\muhighimp{},\,\muclustlowimp{},\,\muunclust{},\,\muclust{},\,\munight{} \}$, 
a superset of the parameters entering $\Lsurf{}$ in Eq.~\ref{eq:surface_likelihood}.

\begin{figure}[tp]
	\includegraphics[width=\columnwidth]{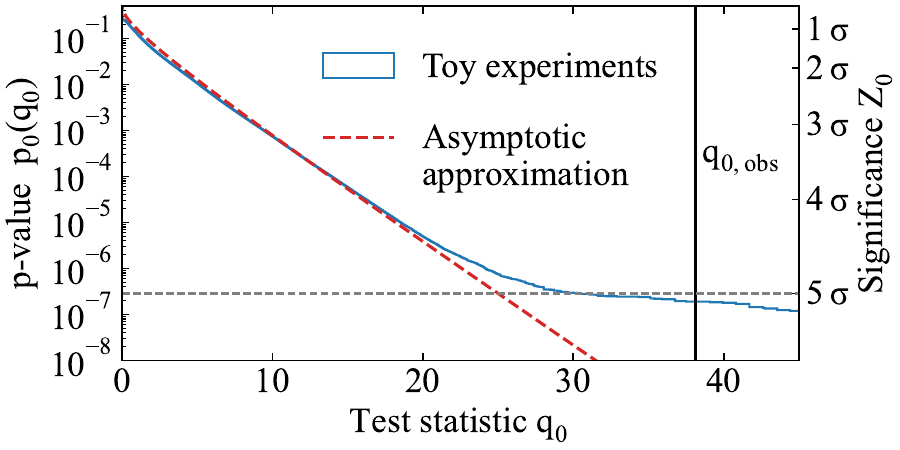}
	\caption{
		Dependence of the $p$-value $p_0$, defined in Eq.~\ref{eq:p-value}, on the test statistic $q_0$
		under the null hypothesis $H_0$.
		The observed test statistic value $q_{0,\mathrm{obs}}$ is indicated by the vertical black line.
		The analytic result for $p_0(q_0)$ from Ref.~\cite{asymptotic_likelihood}, valid in the asymptotic 
		approximation, is shown by the red dashed line.
	}
	\label{fig:test_stat}
\end{figure}

To measure the compatibility of the observed dataset with the background-only hypothesis $H_0$, we consider
the discovery $p$-value \cite{pdg-statistics, asymptotic_likelihood}
\begin{equation}
	p_0(q_{0, \mathrm{obs}}) = \int_{q_{0, \mathrm{obs}}}^{\infty} dq_0 \,\, p(q_0 | H_0),
	\label{eq:p-value}
\end{equation}
calculated in terms of the profile-likelihood test statistic
\begin{equation}
	q_0 = -2 \log \frac{L(\musig = 0, \doublehat{\nuvec}(0))}{L(\musigest, \hat{\nuvec})}.
\end{equation}
The model parameters defining $H_0$ are obtained from a background-only fit, 
i.e.~$H_0: \{\musig = 0, \nuvec = \doublehat{\nuvec}(0)\}$.
The distribution $p(q_0 | H_0)$ appearing in Eq.~\ref{eq:p-value} is obtained from toy experiments, 
shown in Fig.~\ref{fig:test_stat}.
It is compared with the analytic result from Ref.~\cite{asymptotic_likelihood}, valid in the limit of large data-sample
size where the MLEs are distributed according to a normal distribution.
In our case, the presence of small event counts in the likelihood and non-Gaussian MLEs means that the tails of the true test-statistic 
distribution $p(q_0 | H_0)$ deviate from the analytic result and the use of toy experiments for the calculation of 
the $p$-value $p_0$ is important.
The discovery significance corresponding to $p_0$ is calculated as $Z_0 = \Phi^{-1}(1 - p_0)$, 
where $\Phi$ is the CDF of the standard normal distribution.

The usage of the profile-likelihood test statistic $q_0$ in Eq.~\ref{eq:p-value} is advantageous because the
distribution $p(q_0 | \nuvec)$ is approximately pivotal \cite{wilks}, i.e.~the inferred $p$-value and significance
depend only weakly on the (unknown) true values of the nuisance parameters $\nuvec$.
This is not generally the case for other choices of the test statistic, e.g.~the distribution of the 
observed event count in the passing region depends on the nuisance parameters.

\begin{figure}[b]
	\includegraphics[width=\columnwidth]{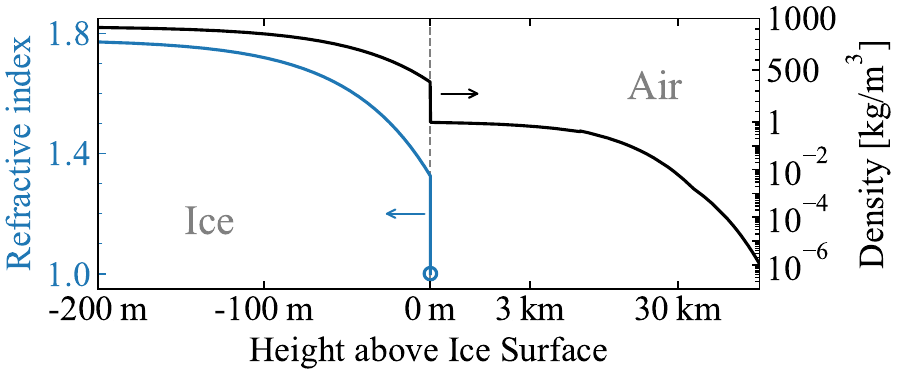}
	\caption{
		Material density (black, right scale) and ice refractive index profile (blue, left scale) 
		used in the microscopic shower simulation.
	}
	\label{fig:material}
\end{figure}

\paragraph{Stability tests}
Several tests have been performed to verify that the results given in the main text are stable against modifications
of the background estimation procedure.
The significance remains above 5\,$\sigma$ if the fit range used for the thermal background estimate is varied
in the region achieving good fit quality, or if
the gamma constraint term for $\tautir{}$ is replaced with a uniform constraint in the interval $[0, 2 \times 10^{-2}]$.
This indicates that both background components are subdominant relative to the on-surface background.

\section{Simulation}
\subsection{Microscopic impacting-shower simulation}
\noindent
We next present details on the simulations used throughout this analysis.
Fig.~\ref{fig:material} shows the material density and refractive index distributions used in the microscopic
simulations.
The \texttt{SouthPoleJul} atmosphere model in \corsika{} is used, which describes the atmosphere as a 5-layer material
with parameters corresponding to conditions at the South Pole in July.
The ice is modeled through the single-exponential density profile 
$\rho(z) = 0.923\,\mathrm{g/cm}^3 \left[1 - 0.582 \exp\left(0.0202 \frac{z}{1\,\mathrm{m}}\right) \right]$,
where $z$ is the vertical coordinate and $z=0$ at the surface.
This choice is related to the refractive index profile (cf.~Eq.~1 in Ref.~\cite{ara-lowthresh-analysis}) through the
phenomenological relation $n(z) = 1 + \rho(z)\times 0.845\,\mathrm{cm}^3/\mathrm{g}$~\cite{gow-kovacs}.
We use the \sibyll{} \cite{sibyll} high-energy hadronic interaction model, the 
\fluka{} \cite{fluka-general-1, fluka-general-2, fluka-version} 
low-energy hadronic interaction model, and \proposal{} \cite{proposal-1, proposal-2} 
to evolve the electromagnetic cascade.
Thinning is applied to electromagnetic particles falling below a fraction of $10^{-6}$ of the primary energy,
and their kinetic cutoff energy is set to 0.5\,MeV (0.3\,GeV for hadrons and muons).
To calculate the signal observed by the receiver, we use \eisvogel{} to convolve positron and electron
tracks in the cascade with an electrodynamic Green's function \cite{eisvogel, eisvogel_arena}.
The Green's function is obtained via the reciprocity relations of Maxwell's equations \cite{reciprocity}
by placing a point dipole at the receiver location and using the finite-difference time-domain solver MEEP~\cite{meep}
to calculate
the electric field generated by exciting the dipole with a current waveform corresponding to the instrumental impulse response~\cite{PA_instrument_paper}.

\begin{figure}[b]
	\includegraphics[width=\columnwidth]{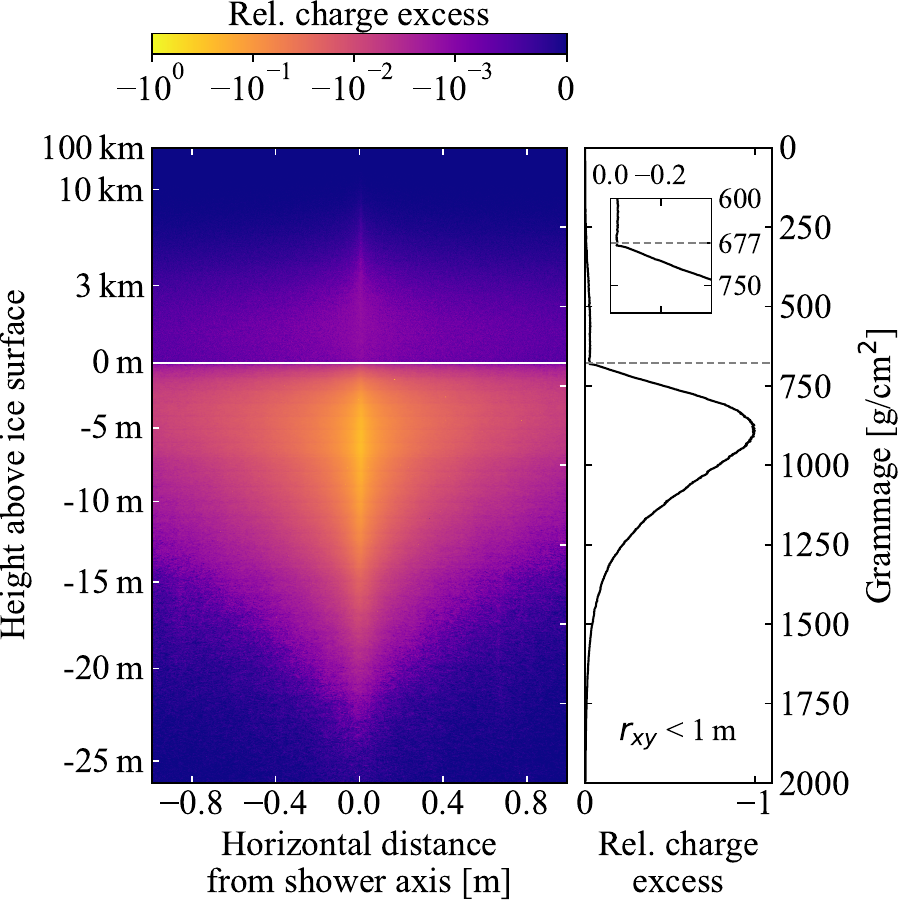}
	\caption{
		Left panel: two-dimensional charge excess distribution normalized to the maximum value
		in a 1\,m-deep slice around the shower axis of a vertical $10^{17}$\,eV proton shower, 
		in bins of constant grammage.
		Right panel: total relative charge excess within 1\,m transverse radius $r_{xy}$ of the shower core 
		as a function of grammage.
	}
	\label{fig:charge_excess}
\end{figure}

\paragraph{Charge excess evolution and transition radiation}
Fig.~\ref{fig:charge_excess} shows the charge-excess distribution, normalized to the maximum charge excess,
within 1\,m of the shower axis for a $10^{17}$\,eV vertically-impacting air shower.
The charge excess $\Delta q$ is defined as $\Delta q = n^{+} - n^{-}$, where $n^{+}$ and $n^{-}$ are the numbers
of positrons and electrons in a bin, respectively.
At impact, the dense shower core is dominated by high-energy photons and shows a charge excess of order 5\%
of the maximum in-ice charge excess.
Transition radiation (TR) produced at impact is expected to be emitted in the forward direction
close to the shower axis and near the Cherenkov angle \cite{Original_faerie}, where it can interfere with the 
Askaryan component.
Owing to the relatively-small charge excess at impact, TR is expected to be a subleading emission mechanism compared
to the Askaryan effect in ice.
This is indeed confirmed by our simulation, which shows that the intensity of the radiation emitted in the direction expected 
for forward-TR is smaller by a factor of $>10^4$ compared to the intensity at the Cherenkov angle.
(Note that the analytical calculations of Ref.~\cite{Original_faerie} predict a large TR component
only due to the inclusion in the charge-excess calculation of particles outside the coherently-emitting shower core.)

\subsection{Monte Carlo simulation}
\noindent
To efficiently generate a large sample of impacting-CR events,
we use a fast parametrized neutrino model that includes both charged-current and neutral-current
interactions, following the general approach of Ref.~\cite{coleman2024iniceaskaryanemissionair}.
Their simulation considered a narrow trigger band with a bandwidth of $\approx$ 100\,MHz, within which
the signals obtained with the parametrized neutrino shower and a microscopic simulation 
of an impacting CR shower core are very similar in shape.
In this scenario, the relative amplitude scale factor $f$ is then uniquely defined as the scale factor of the time-domain
signals.
The ARA trigger uses a much wider band of 150\,MHz--720\,MHz, 
which can lead to differences in the signal shape between the parametrization and the microscopic simulation.
We account for the resulting ambiguity by defining two quantities \flin{} and \fRMS{}, which act as scale factors
of the coherently-added and incoherently-added Fourier amplitudes, respectively,
\begin{align}
	\flin{} &= \frac{\int d\omega \, |V_{\mathrm{C8}}(\omega)|}{\int d\omega \, |V_{\mathrm{param.}}(\omega)|},\\
	\fRMS{} &= \frac{\sqrt{\int d\omega \, |V_{\mathrm{C8}}(\omega)|^2}}{\sqrt{\int d\omega \, |V_{\mathrm{param.}}(\omega)|^2}}.
\end{align}
In these expressions, $V_{\mathrm{C8}}$ and $V_{\mathrm{param.}}$ are the signals simulated by the microscopic \corsika{} approach
and the parametrized simulation, respectively, including the in-borehole antenna response and instrument response.

\begin{figure}[t]
	\includegraphics[width=\columnwidth]{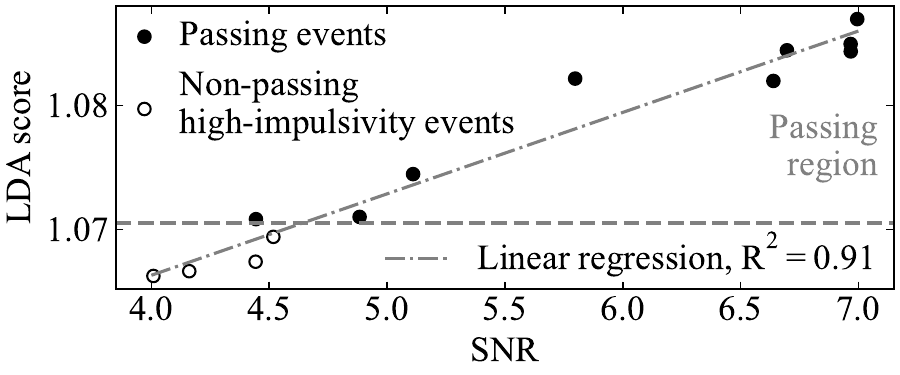}
	\caption{Relation between LDA score and SNR for high-impulsivity ($\imp{} > 25$) events in the zenith region of interest
		above or near the analysis selection cut (dashed line).
		The dash-dotted line shows the best-fit linear relationship between the two variables, with a coefficient of 
		determination of $0.91$.}
	\label{fig:lda_snr}
\end{figure}

\begin{figure}[b]
	\includegraphics[width=\columnwidth]{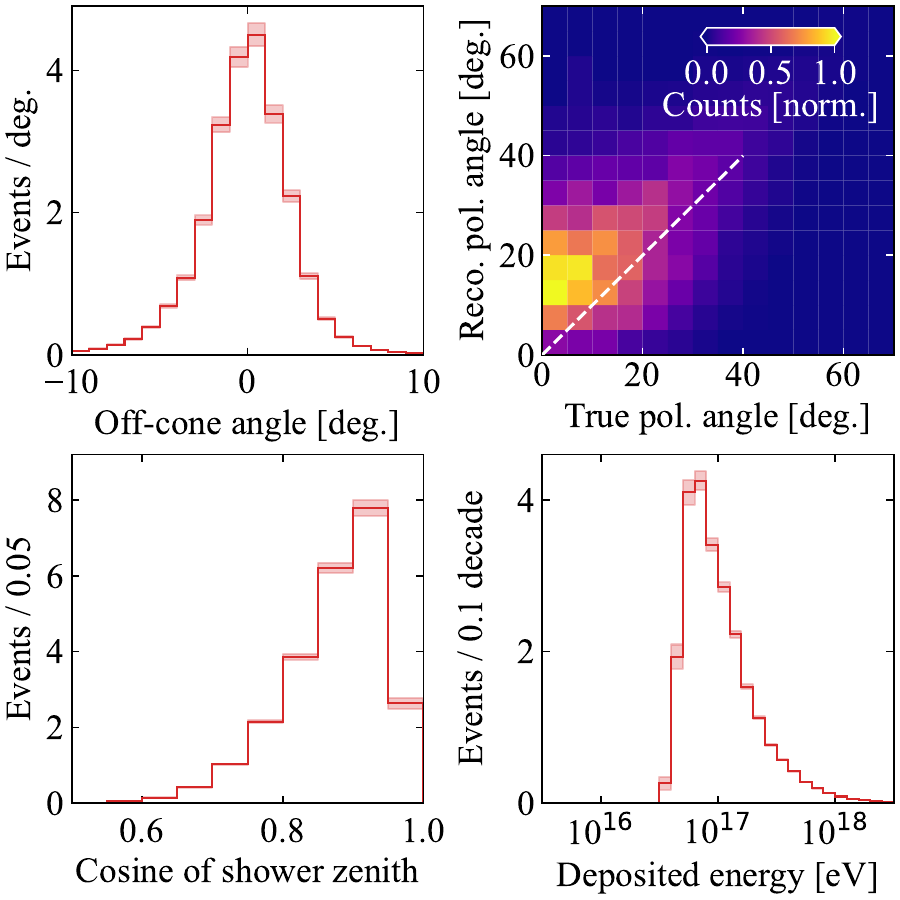}
	\caption{
		Top left: simulated distribution of the off-cone angle, defined as the signed difference between
		the Cherenkov angle and the view angle at shower maximum, both measured relative to the shower axis. 
		Top right: reconstructed polarization angle (cf.~Fig.~4) compared with the true polarization angle, 
		with color indicating the bin content.
		Bottom left: simulated distribution of the CR shower axis zenith angle. 
		Bottom right: simulated distribution of in-ice deposited energy.
		All distributions are normalized to the event yield predicted by the simulation for $f = 0.6$.
	}
	\label{fig:mc}
\end{figure}

\paragraph{Simulated analysis threshold}
To simulate the analysis selection, we apply a requirement on the simulated event SNR, equivalent to the LDA selection cut 
used in Ref.~\cite{ara-lowthresh-analysis}.
In constructing this cut, we observe that the LDA score for passing near-threshold events is approximately linearly related to SNR,
cf.~Fig.~\ref{fig:lda_snr}.
This is the result of the LDA score being defined as a linear combination of its input variables, the most discriminating 
of which are themselves strongly correlated with SNR.
The derived SNR threshold of 4.6 corresponds to a trigger efficiency of $\approx 100\%$ \cite{PA_instrument_paper}; 
trigger effects therefore do not play a role in the event selection.

\begin{figure}[t]
	\includegraphics[width=\columnwidth]{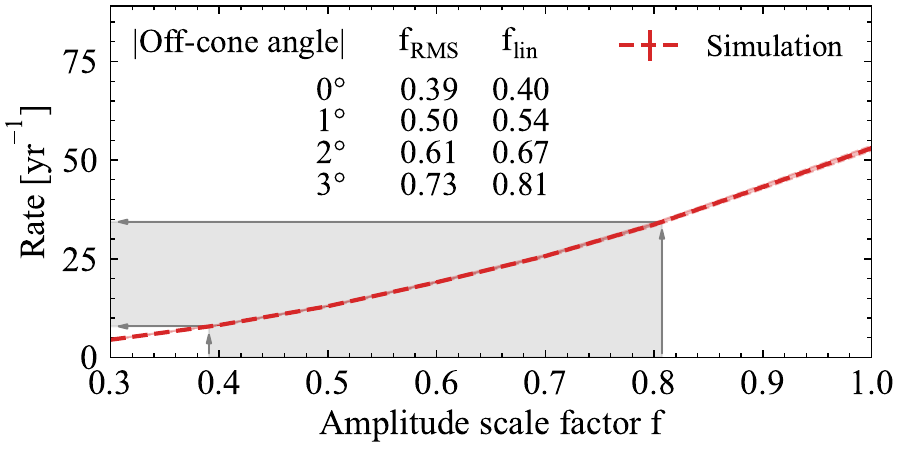}
	\caption{
		The red dashed line shows the simulated rate of in-ice Askaryan events from impacting CR shower cores
		in the zenith region of interest as a function of the phenomenological emission scale
		factor $f$.
		The table gives the numerical values for the scale factors $\fRMS$ and $\flin$ at (absolute) off-cone angles of 
		up to $3^\circ$.
		Their range (shown in gray) determines the event rate interval quoted in the main text.
	}
	\label{fig:rate}
\end{figure}

\paragraph{Simulated distributions and event rate}
Fig.~\ref{fig:mc} shows several example simulated distributions.
The evolution of the simulated passing event rate with the scale factor $f$ is shown in Fig.~\ref{fig:rate}.
As shown by the table inset, $f$ depends strongly on the off-cone angle at which the event triggers.
To construct an interval for the simulated event rate, we therefore take the interval spanned by \flin{} and \fRMS{}
for an off-cone angle of up to $3^\circ$.
This is the angle range relevant for our event selection; around 80\% of simulated passing events are located
within $3^\circ$ of the Cherenkov angle, independent of $f$.
This range of $f$ is then translated into an event rate interval, cf.~Fig.~\ref{fig:rate}.

\section{Test for a shower-like source}
\subsection{Derivation of Eq.~2}
\noindent
Here we derive in more detail Eq.~2 for the case of a Gaussian charge excess profile, given by
\begin{equation}
	q(z) = q_0 \exp \left(-\frac{1}{2}\frac{z^2}{L^2}\right).
\label{eq:charge_profile}
\end{equation}
For the in-ice cascade induced by an impacting CR shower core, the shower length scale is 
$L \approx 5\,\mathrm{m}$ (cf.~Fig.~\ref{fig:charge_excess}), while typical observation wavelengths
are $\lambda\approx1$\,m, such that $\lambda / L \ll 1$.
To calculate the angular and frequency dependence of the electric field in this case, inserting Eq.~\ref{eq:charge_profile} into Eq.~1,
\begin{equation}
  ||\mathbf{E}(\om, \theta)|| \propto i\om \sin\theta \int dz\, q(z) \exp \left(i z \frac{n \om}{c} \left(\cos\theta_c-\cos\theta\right)\right),
  \label{eq:emission_shower}
\end{equation}
and evaluating the Fourier transform returns
\begin{equation}
	||\mathbf{E}(\om, \theta)|| 
	\propto i \frac{L}{\lambda} \sin\theta \exp\left(-2\pi^2 \left(\frac{\cos\theta_c - \cos\theta}{\lambda/L}\right)^2\right),
	\label{eq:askaryan_cone}
\end{equation}
written in terms of the in-ice wavelength $\lambda = 2\pi c / (n \omega)$ and the view angle $\theta$.
Using this expression to explicitly calculate the logarithmic intensity ratio from Eq.~2
and expanding in powers of $\lambda_{1,2}/L$, we find for $\omega_2 > \omega_1$,
\begin{multline}
	\frac{d\log ||\mathbf{E}(\om_2, \theta)||^2}{d\theta} \biggr/ \frac{d\log ||\mathbf{E}(\om_1, \theta)||^2}{d\theta} = \\
	= \left(\frac{\omega_2}{\omega_1}\right)^2 \left[1 - \frac{1}{4\pi^2}\frac{\lambda_1^2 - \lambda_2^2}{L^2} \frac{\cos\theta}{\sin^2\theta\,(\cos\theta-\cos\theta_c)} + \right. \\
	\left. +\,\mathcal{O}\left(\frac{\lambda_1^2}{L^2}\frac{\lambda_1^2 - \lambda_2^2}{L^2}\right)\right].
  \label{eq:hflf_slope}
\end{multline}
The $\om_2^2/\om_1^2$ frequency ratio in the leading-order term is responsible for the result in Eq.~2, while
the next-order term is subdominant except close to the Cherenkov angle $\theta_c$.
Its deviation from unity is below 60\% for \mbox{$|\theta-\theta_c| > 0.5^\circ$} and below 20\% for $|\theta-\theta_c| > 1^\circ$ 
when $f_1 = 170$\,MHz, $f_2 = 285$\,MHz.
Eq.~\ref{eq:hflf_slope} implies that intensity falls off faster as a function of off-cone angle at higher frequencies.

This diffractive property of the Askaryan signal is in contrast with electrically small transmitters, where $\lambda_{1,2}/L > 1$ 
and the dependence on angle and frequency approximately factorizes, such that $||\mathbf{E}(\om, \theta)||^2 = f(\omega)g(\theta)$
for some (transmitter-dependent) functions $f$ and $g$.
Then $\log ||\mathbf{E}(\om, \theta)||^2 = \log f(\omega) + \log g(\theta)$ and we get
\begin{equation}
  \frac{d\log ||\mathbf{E}(\om_2, \theta)||^2}{d\theta} \biggr/ \frac{d\log ||\mathbf{E}(\om_1, \theta)||^2}{d\theta} =
  1.
  \label{eq:pointlike}
\end{equation}
This expresses the fact that, when diffraction effects are absent, signal power is lost at identical rates across different frequencies.
This is also visible from Eq.~\ref{eq:askaryan_cone} directly: if $\lambda/L \gg 1$, then the exponential term is slowly varying and 
Eq.~\ref{eq:pointlike} approximately holds.

\begin{table}
	\centering
	\renewcommand{\arraystretch}{1.2}
	\begin{tabular}{r|cc}
		& LF band & HF band \\
		& $f = 170$\,MHz & $f = 285\,$MHz \\ \midrule
		$\lambda / L$ & 0.26 & 0.16 \\
		$\ipredmax/\ipred$, $\Delta\theta = 3^\circ$ & 1.9 & 7.5 \\
		$\ipredmax/\ipred$, $\Delta\theta = 5^\circ$ & 7.4 & 380 \\
		$\ipredmax/\ipred$, $\Delta\theta = 7^\circ$ & 65 & $2\times10^5$
	\end{tabular}
	\caption{
		Numerical values for the intensity ratios $\ipredmax/\ipred$ at different off-cone angles 
		$\Delta\theta = \theta - \theta_c$, calculated using Eq.~\ref{eq:askaryan_cone}.
		The selected frequencies $f$ correspond to the lower band edges of the chosen LF and HF bands, and the
		chosen view angles are representative of the A5 detector geometry.
		A value of $n = 1.35$ is used in the calculation, corresponding to the refractive index of near-surface
		glacial ice.
	}
	\label{tab:hflf}
\end{table}

It is instructive to evaluate the above equations for typical frequencies in the LF and HF bands.
Table \ref{tab:hflf} shows the intensity ratio $\ipredmax/\ipred = ||\mathbf{E}(\theta_c)||^2 / ||\mathbf{E}(\theta)||^2$
at different off-cone angles $\Delta\theta = \theta - \theta_c$.
The Cherenkov cone is sufficiently wide to generate an intensity ratio in the LF band of order 10 when
illuminating receivers across $2\Delta\theta\approx10^\circ$.
This is comparable to the angular scale instrumented by A5 and thus sufficient for an event to be visible across several A5 receivers.
Conversely, the intensity ratio observed in the HF band by the same receivers is of order 100, comparable to the usable 
dynamic range of the A5 instrument.
This combination of instrumented bandwidth, dynamic range, and geometric arrangement of the A5 receivers allows the diffractive
nature of the Askaryan signature to be identified, provided that sufficiently many receivers are illuminated.

\subsection{Additional information for Fig.~5}
\noindent
For the results of Fig.~5, the signal intensity is calculated in the bands 170\,MHz--285\,MHz (``LF'')
and 285\,MHz--400\,MHz (``HF'').
The lower band edge of the LF band is chosen to be above the cutoff-frequency of the highpass filter defining
the overall detector band, ensuring a stable amplitude response.
The upper edge of the HF band is selected to avoid the notch filter at 450\,MHz and to ensure a predominantly 
single-modal antenna response.
Both bands are chosen to be adjacent and have the same bandwidth.
As a result, the finite frequency resolution generates a small (around 5\%) positive Pearson correlation between the 
intensities measured in the two bands, which is subdominant to the statistical and systematic uncertainties and
therefore not included in the likelihood of Eq.~5.
No optimization of the relative bandwidth has been performed; we note that selecting the bands based on a large 
ensemble of simulated impacting-CR events may enhance the statistical power of the discriminant.

\begin{figure}[t]
	\includegraphics[width=\columnwidth]{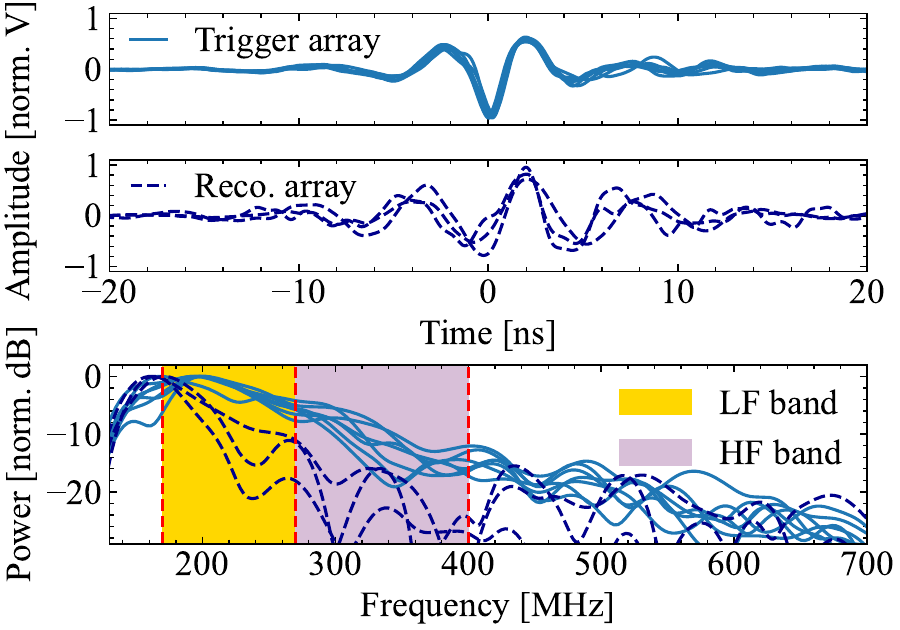}
	\caption{
		Top, center: Time domain VPol waveforms for the passing event of Fig.~5 with the instrumental phase response removed, 
		normalized to their peak-to-peak amplitudes and windowed around the peak using a 25\,ns-wide Tukey window. 
		The traces shown here correspond to the channels included in Fig.~5.
		Signals observed in the trigger array (reconstruction array) are shown in light blue (dark blue).
		Bottom: Corresponding relative power density spectra, normalized to maximum power density.
		The HF and LF bands used for the calculation of the discriminant are highlighted.
	}
	\label{fig:spectrum_plot}
\end{figure}

Fig.~\ref{fig:spectrum_plot} shows the time-domain waveforms and frequency spectra observed in different A5 
receivers for the passing event of Fig.~5. 
The triggering phased-array channels have larger signal amplitudes and relatively more high frequency content, 
while the signal in the reconstruction array is dominated by power in the LF band.
This loss of signal intensity at high frequencies is characteristic of an off-cone Askaryan signal.

\begin{figure}[t]
	\includegraphics[width=\columnwidth]{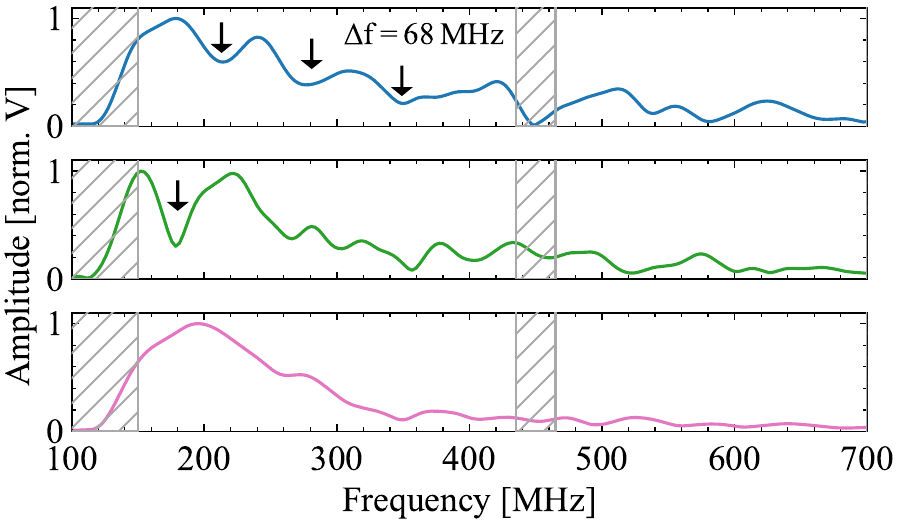}
	\caption{
		Relative power density spectra, obtained with 35\,ns-wide Tukey windows, 
		which show examples of the features mentioned in the text. 
		Top, center: Single-channel spectra with visible interference fringes,
		both of which are from passing events with $\Delta\log L < 0$. 
		Bottom: Single-channel spectrum with no evidence of interference fringes,
		taken from the high-SNR passing event of Fig.~5, which has $\Delta\log L > 0$.
	}
	\label{fig:fringes}
\end{figure}

It is interesting to point out that several passing events show spectral dips resembling interference 
fringes (visible in Fig.~2, but cf.~Fig.~\ref{fig:fringes} for details).
The observed fringe spacing is 
$\approx70\,\mathrm{MHz}$, corresponding to a time delay between two interfering
signal components of $\approx14\,$ns.
Simulation shows that this is consistent with multipath propagation of in-ice Askaryan radiation from an inclined
impacting air shower core, where one component propagates directly to the receiver while the
other undergoes a reflection at the ice-air interface \cite{c8-reflection}.
Near-surface glacial ice further exhibits known density fluctuations with a scale length of the order of 
1\,m \cite{cfm}, potentially leading to the (partial) reflection of certain spectral components akin to a distributed 
Bragg reflector \cite{stratified-reflection}.
Future full-electrodynamics simulation efforts \cite{eisvogel} accounting for wavelength-scale material features 
will quantify these propagation effects.
Here, we only note that they appear to be present in our dataset at some level, inducing channel-to-channel
deviations from the smooth dependency of Eq.~\ref{eq:hflf_slope} and potentially 
interfering with the intensity scaling assumed in Fig.~5.
 
\end{document}